\documentclass[pra,aps,10pt,twocolumn,showpacs,amsmath,amssymb]{revtex4-2}

\usepackage[english]{babel}
\usepackage{afterpage}
\usepackage{graphicx}
\usepackage{dcolumn}
\usepackage{bm}
\usepackage{color}
\usepackage{slashed}
\usepackage{latexsym}

\newcommand{\dd}{\mathrm{d}}  
\newcommand{\ii}{\mathrm{i}}  
\newcommand{\ee}{\mathrm{e}}  

\newcommand{\me}{m_{\mathrm{e}}}  
\newcommand{\bbar}[1]{\check{#1}}  %

\begin{document}

\title{Trident pair creation by a train of laser pulses: \\ Resonance, threshold, and carrier envelope phase effects}
\author{J. Z. Kami\'nski}
\email[E-mail address:\;]{jkam@fuw.edu.pl}
\author{K. Krajewska}
\email[E-mail address:\;]{kkraj@fuw.edu.pl}
\affiliation{Institute of Theoretical Physics, Faculty of Physics, University of Warsaw, Pasteura 5,
02-093 Warsaw, Poland}
\date{\today}

\begin{abstract}
General formulation in the realm of strong-field quantum electrodynamics is provided for a process that occurs in the presence of a train of laser pulses
and, in the tree level, is represented by a two-vertex Feynman diagram with exchange of a virtual photon. A scheme of retrieving resonances
in the corresponding probability distributions is also formulated in these general settings. While the presented formalism is applicable 
to a variety of processes like electron-positron pair creation and annihilation, M\"oller scattering, 
Bhabha scattering, etc., we illustrate it for a trident process. Specifically, we consider electron-positron pair creation in the muon--laser-field 
collisions. We demonstrate that the probability distributions exhibit integrable singularities close to the threshold of pair creation. Also,
a variety of resonances is observed that originate from the poles of the Feynman photon propagator. While those resonances are, in general, obscured
by strong quantum interferences, we show that they can be isolated by changing the carrier envelope phase of the driving laser pulses. In that case,
while transformed into the Lorentz-Breit-Wigner shape profile, the resonance position and width can be determined.
\end{abstract}


\maketitle

\section{Introduction}
\label{sec::intro}

Verifying predictions of strong-field quantum electrodynamics (QED) requires unprecedented electromagnetic fields of the order of the critical 
Sauter-Schwinger field, ${\cal E}_S\approx 1.3\times 10^{18}$~V/m (see, reviews~\cite{MTB2006,EKK2009,PMHK2012,TitovRev,HartinRev,Gonoskov,Fedotovnew} on the development of 
strong-field QED). This presents a significant experimental challenge. However, the required field strength regime can be approached
in the rest frame of a relativistic particle interacting with a high-intensity laser field. This idea was already exploited in the pioneering 
SLAC E144 experiment, in which a two-step trident process was realized~\cite{Stanford1,Stanford2}. In that experiment, a high-energy photon was first produced
in the electron--laser-beam collisions. Next, the photon was backscattered by the electron beam to produce electron-positron ($e^-e^+$) pairs.
Note that in this experiment the signal of produced pairs was largely suppressed, following a typical perturbative scaling with the laser field strength.
The latter equals $\mu=|eA_0|/(m_{\rm e}c)$ where $e$ is the electron charge, 
$m_{\rm e}$ the electron mass, $c$ the speed of light, and $A_0$ is the amplitude of the vector potential describing the laser field.
Thus, the aforementioned experiment covered the parameter region for which $\mu<1$.

With more powerful lasers available today, the community is in the position now to perform next-generation experiments such as upcoming LUXE
(Laser Und XFEL Experiment; see Refs.~\cite{Luxe1,Luxe2}) at the European XFEL and the E320 at FACET-II (Refs.~\cite{Facet1,Facet2}).
They will aim at probing the strong-field QED effects in a nonperturbative regime, i.e., for $\mu>1$. This includes a one-step trident pair production
which occurs with the emission of a virtual, instead of a real, photon. Specifically, at $\mu\gg 1$, the process becomes sizable
if another parameter, known as the quantum parameter, $\chi_{\rm Ritus}=\mu\hbar\omega_{\rm L}/(m_{\rm e}c^2)$, with the reduced Planck
constant $\hbar$ and the carrier laser frequency $\omega_{\rm L}$, becomes larger than 1. This is exactly the parameter region promised by the new experiments.

Regarding a trident pair production, a number of theoretical works have been 
published in the past~\cite{HMK2010,tridentKK2010,kkraj2011,tridentAnton2011,tridentKK2015,tridentPRA89,Acosta2019,King1,King2,Dinu1,Dinu2,Mackenroth,Dinu3,Torg1,Torg2}.
They study an impact of electromagnetic fields on the trident process, which includes either constant crossed fields~\cite{King1,King2}
or propagating plane waves in various forms (i.e., monochromatic~\cite{HMK2010,tridentKK2010,kkraj2011,tridentAnton2011,tridentPRA89}, 
modulated~\cite{tridentKK2015}, or pulsed plane waves~\cite{Dinu1,Dinu2,Dinu3,Torg1,Torg2,Acosta2019,Mackenroth}). 
Among other topics, the conditions for which either the one-step or the two-step scenario is dominant have been investigated. Along with these
recent results, new methods of calculating the one-step trident process have been introduced; namely, ressumation methods (see, e.g., Refs.~\cite{Torg1,Torg2}). 
Much attention has been also paid to divergences arising in the probability amplitude of the trident process~\cite{HMK2010,tridentKK2010,kkraj2011,tridentAnton2011}. They originate 
from the poles of the Feynman photon propagator and are due to the infinite spatiotemporal extent of the driving field; we shall refer to them
as {\it Ole\u{\i}nik resonances}~\cite{Olejnik1967,OlejnikBook,OlejnikRes1,OlejnikRes2,rosh1,rosh2,rosh3,rosh4,PanekMoller2004,PanekMoller2004a,AtomsFelipe2019}. In order
to treat them, the regularized propagator approach was introduced in Ref.~\cite{HMK2010}. Moreover, it has been shown in Ref.~\cite{tridentAnton2011} 
that divergences are absent provided that finite size effects of the pulsed laser fields are accounted for. These developments indicate that there is
still work to be done regarding the theoretical description of the trident pair production. In fact, the same concerns other
two-vertex processes that occur with exchange of a virtual photon such as electron-positron annihilation, M\"oller scattering or Bhabha scattering.

The aim of the current paper is twofold. First of all, we provide a general formulation of a QED process that occurs in a presence of an infinite train of
laser pulses, being described in the first leading order by a two-vertex Feynman diagram with exchange of a virtual photon. Second, we further
advance the theoretical understanding of the trident process. Namely, we demonstrate that the probability rates of trident pair creation exhibit 
integrable singularities at the thresholds. Another aspect of our investigations concerns resonances. Modeling numerically an infinite train of pulses
as very long but finite, we observe rather complex resonant structures in the energy distributions of created particles. Interestingly, one can resolve 
individual resonances in the spectra by adjusting the carrier envelope phase of the driving pulses. In that case, they acquire nearly 
Lorentz-Breit-Wigner shape profiles which allow to determine the resonance position and width. Even more, one can also enhance the resonant contribution
to the process by changing the carrier envelope phase. Thus, we propose a new way of phase control
for the trident process. 

The paper is organized as follows. In Sec.~\ref{sec1}, we calculate the strong-field QED probability amplitude for a process described by 
a two-vertex Feynman diagram with an exchange of a virtual photon. In Sec.~\ref{sec::laser} we specify a laser field, whereas
in Sec.~\ref{sec::Volkov} we derive the Volkov solutions of the respective Dirac equation. The Dirac-Volkov current is defined
in Sec.~\ref{sec::current} and the probability amplitude for the process is derived in Sec.~\ref{ampli}. Its 
divergences are analyzed in Sec.~\ref{generalities}, along with a prescription of how to regularize the Feynman photon propagator.
This general formulation is applied next to study the trident process in the laser-field--muon collisions (Sec.~\ref{trid}).
Here, we start by defining the probability distributions for the trident process (Sec.~\ref{sec::distributions}). The appearance of resonances
is demonstrated in Secs.~\ref{sec::oleinik} and~\ref{sec::CEP}. The latter shows also the behavior of the probability distributions
near the pair creation threshold. The sensitivity of the probability distributions to the carrier-envelope phase of the laser field
is shown in Sec.~\ref{sec::CEP}. Then, we demonstrate how it can be used to steer the resonant behavior of the probability spectra
of the trident process (Sec.~\ref{sec::Lorentz}). The concluding remarks are given in Sec.~\ref{conclusions}.

Since now on, in all formulas we shall put $\hbar=1$, meaning that the fine structure constant equals $\alpha=e^2/(4\pi\varepsilon_0c)$, 
where $\varepsilon_0$ is the vacuum permittivity. Our numerical results, on the other hand, shall be presented in relativistic units such that
$c=m_{\rm e}=\hbar=1$. Also, we will use the Einstein summation convention with the metric signature $(1,-1,-1,-1)$.

\section{General theoretical formulation}
\label{sec1}

In this section, we provide a general framework to describe an arbitrary strong-field QED process, that in the lowest order of perturbation theory
is represented by a two-vertex Feynman diagram with an exchange of a virtual photon. Our formulation is valid when the process is accompanied by
an infinite train of identical laser pulses, as explained in the next section.

\subsection{Laser field}
\label{sec::laser}

We describe a laser field by an electromagnetic potential,
\begin{equation}
A(x)\equiv A(k\cdot x)=A_0[\varepsilon_1 f_1(k\cdot x)+\varepsilon_2 f_2(k\cdot x)],
\label{lp1}
\end{equation}
where $A_0$ determines the intensity of the field, whereas the wave four-vector is $k=k^0n\equiv k^0(1,\bm{n})$ with the unit vector $\bm{n}$ pointing in 
the direction of the laser beam propagation (hence, $k\cdot k=0$). The two real four-vectors $\varepsilon_j$, $j=1,2$, normalized such that 
$\varepsilon_j\cdot\varepsilon_{j'}=-\delta_{jj'}$, define the polarization properties of the radiation field, meaning that $k\cdot\varepsilon_j=0$. 
In the following, we also assume that the polarization four-vectors have only space components, i.e., $\varepsilon_j=(0,\bm{\varepsilon}_j)$.
The two real functions $f_j(k\cdot x)$, $j=1,2$, describe the pulsed properties of the laser field. 
While in principle they can be arbitrary, asymptotically they should acquire the same values,
\begin{equation}
\lim\limits_{k\cdot x\rightarrow \pm\infty} f_j(k\cdot x)=f_j^\infty, \quad j=1,2.
\label{lp2}
\end{equation}
Because the electric and magnetic fields do not depend on the constants $f_j^\infty$, without loosing generality, we assume in the following that 
$f_j^\infty=0$ for $j=1,2$.

In this paper, we assume that the light field is periodic in time (with period $T_p$) at any point in space $\bm{x}$. In other words, we consider an infinite
train of pulses. Hence, by defining the fundamental 
frequency of field oscillations $\omega=ck^0=2\pi/T_p$, we can choose the four-vector potential such that
\begin{equation}
A(k\cdot x)=A(k\cdot x+2\pi).
\label{lp3}
\end{equation}
Specifically, for a pulse from the train lasting for time $T_p$,
\begin{equation}
A(0)=A(2\pi)=0.
\label{lp4}
\end{equation}
For our further purpose, we define the pulse shape averages such that 
\begin{equation}
\langle f_j^n\rangle=\frac{1}{2\pi}\int_0^{2\pi}\dd\phi\, [f_j(\phi)]^n,\quad n\in{\mathbb{N}}.
\label{lp5}
\end{equation}
Consequently, we have
\begin{equation}
\langle A\rangle =A_0(\langle f_1\rangle\varepsilon_1+\langle f_2\rangle\varepsilon_2)
\label{lp7}
\end{equation}
and
\begin{equation}
\langle A^2\rangle=\langle A\cdot A\rangle=-A_0^2(\langle f_1^2\rangle + \langle f_2^2\rangle).
\label{lp8}
\end{equation}
For completeness, we note that the electric field component is defined as
\begin{equation}
\bm{\mathcal{E}}(\phi)=-\partial_t\bm{A}(\phi)=-A_0\omega \bigl[\bm{\varepsilon}_1f'_1(\phi)+\bm{\varepsilon}_2f'_2(\phi)\bigr],
\label{lp9}
\end{equation}
where the '\textit{prime}' means the derivative with respect to the phase $\phi=k\cdot x$, whereas the magnetic field becomes 
$\bm{\mathcal{B}}(\phi)=\bm{n}\times \bm{\mathcal{E}}(\phi)/c$. Note that, for both these vectors, the integral over the phase $\phi$ from 0 to $2\pi$ vanishes.

As an example, we consider an infinite train of linearly polarized laser pulses, with shapes defined by the following master function,
\begin{equation}
F(\phi)=\begin{cases}
\mathcal{N}\sin^2\left(\frac{\phi}{2}\right)\sin(N_{\mathrm{osc}}\phi+\chi), & 0\leqslant \phi \leqslant 2\pi , \cr
0 , & \textrm{otherwise}.
\end{cases}
\label{lp10}
\end{equation}
Here, $N_{\mathrm{osc}}$ determines the number of cycles in an individual pulse whereas $\chi$ is the carrier envelope phase (CEP). 
The real number $\mathcal{N}$ is adapted according to the normalization condition chosen for the laser field~\cite{cep1,cep2,BW5} and it depends on the experimental 
conditions. For $N_{\mathrm{osc}}\geqslant 2$, the master function and its first derivative are continuous. In addition, $F(\phi)$ satisfies
the integral constrain,
\begin{equation}
\int_0^{2\pi} \dd\phi F(\phi)=0.
\label{lp11}
\end{equation}
For the linearly polarized field, we assume that $f_2(\phi)=0=f_2^{\prime}(\phi)$ and define the pulse either as
\begin{equation}
f_1(\phi)=F(\phi),\quad f_1^{\prime}(\phi)=F^{\prime}(\phi),
\label{lp12}
\end{equation}
or
\begin{equation}
f_1(\phi)=-\int_0^{\phi}\dd u F(u),\quad f_1^{\prime}(\phi)=-F(\phi).
\label{lp13}
\end{equation}
In the first case, the function $F(\phi)$ describes the vector potential, $\bm{A}(\phi)=A_0F(\phi)\bm{\varepsilon}_1$, whereas in the second 
case it describes the electric field, $\bm{\mathcal{E}}(\phi)=\omega A_0F(\phi)\bm{\varepsilon}_1$. Here, the fundamental frequency $\omega$ 
is related to the carrier laser frequency $\omega_L$ as $\omega_L=N_{\mathrm{osc}}\omega$ [see, Eq.~\eqref{lp10}]. 
While this model will be considered when performing numerical calculations in Sec.~\ref{trid}, the formulas derived 
in the following sections are for a general vector potential~\eqref{lp1}.

\subsection{Volkov solutions}
\label{sec::Volkov}

Our choice of the electromagnetic potential \eqref{lp1} is motivated by the possibility of constructing the exact solution of the Dirac equation 
for a fermion of the rest mass $m$ and charge $ze$ ($e=-|e|$ and integer $z$),
\begin{equation}
[\gamma\cdot(\ii\partial -zeA(k\cdot x))-m c]\psi(x)=0,
\label{vol1}
\end{equation}
known as the Volkov solution. Its explicit form is
\begin{equation}
\psi_{\bm{p}\lambda}^{(\beta)}(x)=\sqrt{\frac{m c}{Vp^0}}\Bigl[1-\beta\frac{ze\slashed{A}(k\cdot x)\slashed{k}}{2k\cdot p}\Bigr]\ee^{-\ii\beta p\cdot x-\ii \mathcal{W}_{\bm{p}}^{(\beta)}(k\cdot x)}u_{\bm{p}\lambda}^{(\beta)},
\label{vol2}
\end{equation}
where $\beta=\pm$ [$\beta=+$ relates to particles and $\beta=-$ to anti-particles],
\begin{equation}
\mathcal{W}_{\bm{p}}^{(\beta)}(k\cdot x)=\int_0^{k\cdot x}\dd\phi\Bigl[\frac{zeA(\phi)\cdot p}{k\cdot p}-\beta\frac{z^2e^2A^2(\phi)}{2k\cdot p} \Bigr],
\label{vol3}
\end{equation}
and $p^\mu=(p^0,\bm{p})$ with $p^0=\sqrt{\bm{p}^2+(m c)^2}$. The index $\lambda=\pm$ discriminates the spin degrees of freedom and the free-particle 
bi-spinors $u_{\bm{p}\lambda}^{(\beta)}$ fulfill the algebraic equation $(\slashed{p}-\beta m c)u_{\bm{p}\lambda}^{(\beta)}=0$. These bispinors 
are normalized such that $\bar{u}^{(\beta)}_{\bm{p}\lambda}u^{(\beta')}_{\bm{p}\lambda'}=\beta\delta_{\beta\beta'}\delta_{\lambda\lambda'}$ 
(with $\bar{u}^{(\beta)}_{\bm{p}\lambda}=[u^{(\beta)}_{\bm{p}\lambda}]^{\dagger}\gamma^0$ being the Dirac conjugation)
and they
satisfy the completeness relation, self-consistent with the normalization one, 
\begin{equation}
\sum_{\beta=\pm}\sum_{\lambda=\pm}\beta u^{(\beta)}_{\bm{p}\lambda}\bar{u}^{(\beta)}_{\bm{p}\lambda}= I_{4\times 4},
\label{vol4}
\end{equation}
where $I_{4\times 4}$ is the four by four unit matrix. Moreover, the quantization volume $V$ defines the density of fermion states which, if not 
accounting for the spin degrees of freedom, is equal to $V\dd^3p/(2\pi)^3$. Equivalently, we can use the normalization in the form~\cite{ItzyksonZuber},
\begin{equation}
\bar{u}^{(\beta)}_{\bm{p}\lambda}\gamma^0u^{(\beta')}_{\bm{p}\lambda'}=\frac{p^0}{m c}\delta_{\beta\beta'}\delta_{\lambda\lambda'},
\label{vol5}
\end{equation}
that leads to the completeness condition
\begin{equation}
\sum_{\beta=\pm}\sum_{\lambda=\pm} u^{(\beta)}_{\bm{p}\lambda}\bar{u}^{(\beta)}_{\bm{p}\lambda}=\frac{p^0}{m c}\gamma^0.
\label{vol6}
\end{equation}
Related to this is the orthogonality and completeness 
of the Volkov states which have been discussed in Refs.~\cite{Boca2010,Boca2011,Antonino2018,Wang2019}. Let us also note that 
in our numerical analysis we shall use the Dirac representation for the $\gamma$ matrices.

We further define the function $\mathcal{G}_{\bm{p}}^{(\beta)}(\phi)$,
\begin{equation}
\mathcal{W}_{\bm{p}}^{(\beta)}(k\cdot x)=\Bigl[\frac{ze\langle A\rangle\cdot p}{k\cdot p}-\beta\frac{z^2e^2\langle A^2\rangle}{2k\cdot p} \Bigr]k\cdot x+\mathcal{G}_{\bm{p}}^{(\beta)}(k\cdot x),
\label{vol7}
\end{equation}
which due to the properties of the vector potential discussed in Sec.~\ref{sec::laser} satisfies the conditions: 
$\mathcal{G}_{\bm{p}}^{(\beta)}(\phi+2\pi)=\mathcal{G}_{\bm{p}}^{(\beta)}(\phi)$ for a train and 
$\mathcal{G}_{\bm{p}}^{(\beta)}(0)=\mathcal{G}_{\bm{p}}^{(\beta)}(2\pi)=0$ for a pulse. With these definitions the Volkov state can be recast into the form,
\begin{equation}
\psi_{\bm{p}\lambda}^{(\beta)}(x)=\sqrt{\frac{m c}{Vp^0}}\ee^{-\ii \beta\bar{p}\cdot x}\Phi_{\bm{p}}^{(\beta)}(k\cdot x)u^{(\beta)}_{\bm{p}\lambda},
\label{vol8}
\end{equation}
where
\begin{equation}
\Phi_{\bm{p}}^{(\beta)}(k\cdot x)=\Bigl[ 1-\beta\frac{ze\slashed{A}(k\cdot x)\slashed{k}}{2k\cdot p} \Bigr]\ee^{-\ii \mathcal{G}_{\bm{p}}^{(\beta)}(k\cdot x)}
\label{vol9}
\end{equation}
and
\begin{equation}
\bar{p}=p+\Bigl[\beta\frac{ze\langle A\rangle\cdot p}{k\cdot p}-\frac{z^2e^2\langle A^2\rangle}{2k\cdot p} \Bigr]k.
\label{vol10}
\end{equation}
The advantage of this representation is that the function $\Phi_{\bm{p}}^{(\beta)}(\phi)$ is periodic with the period $2\pi$ and equals $I_{4\times 4}$ 
for $\phi= 0$ and $\phi=2\pi$, i.e., at the beginning and at the end of an individual pulse from the train. Because of those properties, 
$\Phi_{\bm{p}}^{(\beta)}(\phi)$ can be uniformly approximated by the Fourier expansion~\cite{Serov2017}. The quantity $\bar{p}$ defined by Eq.~\eqref{vol10}
is called the dressed four-momentum and bears some similarities with the quasimomentum of electrons moving in the solid periodic structures.

Since we use the reduced amplitude of the vector potential, $\mu$, the dressed four-momentum~\eqref{vol10} becomes
\begin{align}
\bar{p}=& p-\beta z\mu\me c\Bigl(\frac{\varepsilon_1\cdot p}{k\cdot p}\langle f_1\rangle+\frac{\varepsilon_2\cdot p}{k\cdot p}\langle f_2\rangle\Bigr)k
 \\
+& (z\mu\me c)^2\frac{\langle f^2_1\rangle+\langle f^2_2\rangle}{2k\cdot p}\, k. \nonumber
\label{gdress1a}
\end{align}
Because $k\cdot p=k\cdot\bar{p}$ and $\varepsilon_j\cdot p=\varepsilon_j\cdot\bar{p}$, this relation can be easily inverted, resulting in
\begin{align}
p=& \bar{p}+\beta z\mu\me c\Bigl(\frac{\varepsilon_1\cdot \bar{p}}{k\cdot \bar{p}}\langle f_1\rangle+\frac{\varepsilon_2\cdot \bar{p}}{k\cdot \bar{p}}\langle f_2\rangle\Bigr)k
 \\
-& (z\mu\me c)^2\frac{\langle f^2_1\rangle+\langle f^2_2\rangle}{2k\cdot \bar{p}}\, k. \nonumber
\label{gdress1b}
\end{align}
In closing this section, we note that the Volkov solution and, hence, also the dressed four-momentum are gauge-dependent. 
Moreover, for nonvanishing $\langle f_1\rangle$ or $\langle f_2\rangle$, the dressed four-momentum is not on the mass shell, 
meaning that $\bar{p}^2$ depends on $\bm{p}$. Both these deficiencies are going to be discussed below.

\subsection{Dirac-Volkov current}
\label{sec::current}

The elements of the fermionic four-currents,
\begin{equation}
[j^{(\beta_2\beta_1)}_{\bm{p}_2\lambda_2;\bm{p}_1\lambda_1}(x)]^\nu=
\bar{\psi}^{(\beta_2)}_{\bm{p}_2\lambda_2}(x)\gamma^\nu\psi^{(\beta_1)}_{\bm{p}_1\lambda_1}(x),
\label{volc1}
\end{equation}
are of fundamental importance for calculating probability amplitudes of QED processes. 
They are gauge-invariant and satisfy the continuity equation,
\begin{equation}
\partial\cdot j^{(\beta_2\beta_1)}_{\bm{p}_2\lambda_2;\bm{p}_1\lambda_1}(x)=0.
\label{volc2}
\end{equation}
Inserting in Eq.~\eqref{volc1} the explicit form of the Volkov solution~\eqref{vol8} and using the definition of the dressed momenta, they can be represented as
\begin{align}
\label{volc3}
[j^{(\beta_2\beta_1)}_{\bm{p}_2\lambda_2;\bm{p}_1\lambda_1}&(x)]^\nu=
\frac{m c}{V\sqrt{p_1^0p_2^0}}[D_{\bm{p}_2\lambda_2;\bm{p}_1\lambda_1}^{(\beta_2\beta_1)}(k\cdot x)]^\nu \\
\times &\exp\bigl[-\ii(\beta_1\bar{p}_1-\beta_2\bar{p}_2)\cdot x -\ii\mathcal{G}_{\bm{p}_2\bm{p}_1}^{(\beta_2\beta_1)}(k\cdot x) \bigr], \nonumber
\end{align}
where
\begin{align}
\label{volc7}
[D_{\bm{p}_2\lambda_2;\bm{p}_1\lambda_1}^{(\beta_2\beta_1)}&(\phi)]^\nu= [D_{\bm{p}_2\lambda_2;\bm{p}_1\lambda_1}^{(\beta_2\beta_1)(0,0)}]^\nu \\
+&[D_{\bm{p}_2\lambda_2;\bm{p}_1\lambda_1}^{(\beta_2\beta_1)(1,0)}]^\nu f_1(\phi) 
+[D_{\bm{p}_2\lambda_2;\bm{p}_1\lambda_1}^{(\beta_2\beta_1)(0,1)}]^\nu f_2(\phi) \nonumber \\ 
+&[D_{\bm{p}_2\lambda_2;\bm{p}_1\lambda_1}^{(\beta_2\beta_1)(1,1)}]^\nu f_1(\phi)f_2(\phi) \nonumber \\
+&[D_{\bm{p}_2\lambda_2;\bm{p}_1\lambda_1}^{(\beta_2\beta_1)(2,0)}]^\nu f_1^2(\phi) 
+[D_{\bm{p}_2\lambda_2;\bm{p}_1\lambda_1}^{(\beta_2\beta_1)(0,2)}]^\nu f_2^2(\phi) \nonumber 
\end{align}
and
\begin{align}
\label{volc8}
\mathcal{G}_{\bm{p}_2\bm{p}_1}^{(\beta_2\beta_1)}(k\cdot x)=&\int_0^{k\cdot x}\dd\phi 
\Bigl[-\me cz\mu \bigl( (f_1(\phi)-\langle f_1\rangle)\varepsilon_1 \nonumber \\
+&(f_2(\phi)-\langle f_2\rangle)\varepsilon_2 \bigr)\cdot\Bigl(\frac{p_1}{k\cdot p_1}-\frac{p_2}{k\cdot p_2}\Bigr) \nonumber \\
 +&\frac{(\me cz\mu)^2}{2}(f_1^2(\phi)-\langle f_1^2\rangle+f_1^2(\phi)-\langle f_2^2\rangle) \nonumber \\
\times &\Bigl(\frac{\beta_1}{k\cdot p_1}-\frac{\beta_2}{k\cdot p_2}\Bigr)\Bigl] .
\end{align}
In addition, the matrix elements introduced in Eq.~\eqref{volc7} have the form,
\begin{align}
\label{volc10}
[D_{\bm{p}_2\lambda_2;\bm{p}_1\lambda_1}^{(\beta_2\beta_1)(0,0)}]^\nu=&\bar{u}^{(\beta_2)}_{\bm{p}_2\lambda_2}\gamma^\nu u^{(\beta_1)}_{\bm{p}_1\lambda_1},  \\
[D_{\bm{p}_2\lambda_2;\bm{p}_1\lambda_1}^{(\beta_2\beta_1)(1,0)}]^\nu=&\frac{\beta_2\me cz\mu}{2n\cdot p_2}\bar{u}^{(\beta_2)}_{\bm{p}_2\lambda_2}\slashed{n}\slashed{\varepsilon}_1\gamma^\nu u^{(\beta_1)}_{\bm{p}_1\lambda_1} \nonumber \\
&+ \frac{\beta_1\me cz\mu}{2n\cdot p_1}\bar{u}^{(\beta_2)}_{\bm{p}_2\lambda_2}\gamma^\nu \slashed{\varepsilon}_1\slashed{n}u^{(\beta_1)}_{\bm{p}_1\lambda_1},  \nonumber \\
[D_{\bm{p}_2\lambda_2;\bm{p}_1\lambda_1}^{(\beta_2\beta_1)(0,1)}]^\nu=&\frac{\beta_2\me cz\mu}{2n\cdot p_2}\bar{u}^{(\beta_2)}_{\bm{p}_2\lambda_2}\slashed{n}\slashed{\varepsilon}_2\gamma^\nu u^{(\beta_1)}_{\bm{p}_1\lambda_1} \nonumber \\
&+ \frac{\beta_1\me cz\mu}{2n\cdot p_1}\bar{u}^{(\beta_2)}_{\bm{p}_2\lambda_2}\gamma^\nu \slashed{\varepsilon}_2\slashed{n}u^{(\beta_1)}_{\bm{p}_1\lambda_1},   \nonumber \\
[D_{\bm{p}_2\lambda_2;\bm{p}_1\lambda_1}^{(\beta_2\beta_1)(1,1)}]^\nu=&\frac{\beta_2\beta_1(\me cz\mu)^2}{4(n\cdot p_2)(n\cdot p_1)}\bigl[\bar{u}^{(\beta_2)}_{\bm{p}_2\lambda_2}\slashed{n}\slashed{\varepsilon}_2\gamma^\nu \slashed{\varepsilon}_1\slashed{n}u^{(\beta_1)}_{\bm{p}_1\lambda_1} \nonumber \\
&+ \bar{u}^{(\beta_2)}_{\bm{p}_2\lambda_2}\slashed{n}\slashed{\varepsilon}_1\gamma^\nu \slashed{\varepsilon}_2\slashed{n}u^{(\beta_1)}_{\bm{p}_1\lambda_1}\bigr],  \nonumber \\
[D_{\bm{p}_2\lambda_2;\bm{p}_1\lambda_1}^{(\beta_2\beta_1)(2,0)}]^\nu=&\frac{\beta_2\beta_1(\me cz\mu)^2}{4(n\cdot p_2)(n\cdot p_1)}\bar{u}^{(\beta_2)}_{\bm{p}_2\lambda_2}\slashed{n}\slashed{\varepsilon}_1\gamma^\nu \slashed{\varepsilon}_1\slashed{n}u^{(\beta_1)}_{\bm{p}_1\lambda_1},  \nonumber \\
[D_{\bm{p}_2\lambda_2;\bm{p}_1\lambda_1}^{(\beta_2\beta_1)(0,2)}]^\nu=&\frac{\beta_2\beta_1(\me cz\mu)^2}{4(n\cdot p_2)(n\cdot p_1)}\bar{u}^{(\beta_2)}_{\bm{p}_2\lambda_2}\slashed{n}\slashed{\varepsilon}_2\gamma^\nu \slashed{\varepsilon}_2\slashed{n}u^{(\beta_1)}_{\bm{p}_1\lambda_1}.  \nonumber 
\end{align}
We note that the Dirac-Volkov current can be efficiently used in calculations due to the Fourier decomposition,
\begin{align}
[f_1(\phi)]^{j_1}[f_2(\phi)]^{j_2}\exp&\bigl[-\ii\mathcal{G}_{\bm{p}_2\bm{p}_1}^{(\beta_2\beta_1)}(\phi)\bigr] \nonumber \\
=&\sum_{N=-\infty}^{\infty}\ee^{-\ii N\phi}G_{\bm{p}_2\bm{p}_1;N}^{(\beta_2\beta_1)(j_1,j_2)},
\label{fd1}
\end{align}
for $j_1,j_2=0,1,2$. This Fourier series is uniformly convergent~\cite{Serov2017}, as the left-hand side acquires the same values 
for $\phi=0$ and $\phi=2\pi$, and is continuous over the interval $[0,2\pi]$. Hence, we obtain that
\begin{align}
[j^{(\beta_2\beta_1)}_{\bm{p}_2\lambda_2;\bm{p}_1\lambda_1}(x)]^\nu=\frac{m c}{V\sqrt{p^0_2p^0_1}}
\sum_{N=-\infty}^{\infty} & \ee^{-\ii(\beta_1\bbar{p}_1-\beta_2\bbar{p}_2+Nk)\cdot x} \nonumber \\
\times & [j^{(\beta_2\beta_1)}_{\bm{p}_2\lambda_2;\bm{p}_1\lambda_1;N}]^\nu,
\label{fd2}
\end{align}
with
\begin{align}
[j^{(\beta_2\beta_1)}_{\bm{p}_2\lambda_2;\bm{p}_1\lambda_1;N}]^\nu=&[j^{(\beta_2\beta_1)(0,0)}_{\bm{p}_2\lambda_2;\bm{p}_1\lambda_1;N}]^\nu +[j^{(\beta_2\beta_1)(0,1)}_{\bm{p}_2\lambda_2;\bm{p}_1\lambda_1;N}]^\nu \nonumber \\
+&[j^{(\beta_2\beta_1)(1,0)}_{\bm{p}_2\lambda_2;\bm{p}_1\lambda_1;N}]^\nu +[j^{(\beta_2\beta_1)(1,1)}_{\bm{p}_2\lambda_2;\bm{p}_1\lambda_1;N}]^\nu \nonumber \\
+&[j^{(\beta_2\beta_1)(0,2)}_{\bm{p}_2\lambda_2;\bm{p}_1\lambda_1;N}]^\nu +[j^{(\beta_2\beta_1)(2,0)}_{\bm{p}_2\lambda_2;\bm{p}_1\lambda_1;N}]^\nu 
\label{fd3}
\end{align}
and
\begin{equation}
[j^{(\beta_2\beta_1)(j_1,j_2)}_{\bm{p}_2\lambda_2;\bm{p}_1\lambda_1;N}]^\nu=
G_{\bm{p}_2\bm{p}_1;N}^{(\beta_2\beta_1)(j_1,j_2)}
[D_{\bm{p}_2\lambda_2;\bm{p}_1\lambda_1}^{(\beta_2\beta_1)(j_1,j_2)}]^\nu .
\label{fd4}
\end{equation}
Eq.~\eqref{fd2} will be used in Sec.~\ref{ampli} when deriving the formula for the probability amplitude of the respective QED process.

In relation to the fermionic four-currents, which are the fundamental building blocks of QED probability amplitudes, let us go back to the problem of
gauge invariance, mentioned in the previous section. Since the dressed momenta, $\bar{p}_1$ and $\bar{p}_2$, appear in the four-current~\eqref{volc3} 
only in the combination $\beta_1\bar{p}_1-\beta_2\bar{p}_2$, therefore, one can redefine the laser-field dressing such that 
$\bar{p}_j\rightarrow\bar{p}_j+\beta_j s$, $j=1,2$, where the four-vector $s$ could, in principle, be arbitrary. This suggests that the momentum 
dressing does not have a unique physical meaning. However, we can use this ambiguity in order to simplify our further analysis. In the following 
we shall choose~\cite{HHIM2012,KK2012b}
\begin{equation}
s=z\mu\me c(\langle f_1\rangle\varepsilon_1+\langle f_2\rangle\varepsilon_2).
\label{dress1}
\end{equation}
Hence, the shifted dressed momentum becomes now,
\begin{align}
\bbar{p}=& p-\beta z\mu\me c\Bigl(\frac{\varepsilon_1\cdot p}{k\cdot p}\langle f_1\rangle+\frac{\varepsilon_2\cdot p}{k\cdot p}\langle f_2\rangle\Bigr)k \nonumber \\
+& (z\mu\me c)^2\frac{\langle f^2_1\rangle+\langle f^2_2\rangle}{2k\cdot p}\, k 
+\beta z\mu\me c(\langle f_1\rangle\varepsilon_1+\langle f_2\rangle\varepsilon_2). \nonumber \\
=& \bar{p}+\beta z\mu\me c(\langle f_1\rangle\varepsilon_1+\langle f_2\rangle\varepsilon_2),
\label{dress1a}
\end{align}
which is invariant with respect to the gauge transformation $\varepsilon_j=\varepsilon^{\Lambda}_j+a_j(k\cdot x) k$ with arbitrary functions 
$a_j(k\cdot x)$. In fact, for constant $a_j$, such gauge-invariance is used as a test of our numerical calculations.
As it follows from Eq.~\eqref{dress1a}, we have $k\cdot p=k\cdot\bbar{p}$ and 
$\varepsilon_j\cdot p=\varepsilon_j\cdot\bbar{p}+\beta\mu\me c\langle f_j\rangle$. Therefore, Eq.~\eqref{dress1a} is equivalent to 
\begin{align}
p=& \bbar{p}+\beta z\mu\me c\Bigl(\frac{\varepsilon_1\cdot \bbar{p}}{k\cdot \bbar{p}}\langle f_1\rangle+\frac{\varepsilon_2\cdot \bbar{p}}{k\cdot \bbar{p}}\langle f_2\rangle\Bigr)k
 \nonumber \\
-& (z\mu\me c)^2\frac{\langle f^2_1\rangle-2\langle f_1\rangle^2+\langle f^2_2\rangle-2\langle f_2\rangle^2}{2k\cdot \bbar{p}}\, k  \nonumber \\
-&\beta z\mu\me c(\langle f_1\rangle\varepsilon_1+\langle f_2\rangle\varepsilon_2) .
\label{dress1b}
\end{align}
It is important to realize that the new dressed momenta of fermions are on the mass shell, with the effective mass independent of momenta $p$, $k$, and polarization 
four-vectors $\varepsilon_j$. Indeed, one can show that
\begin{equation}
\bbar{p}^2=p^2+(z\mu\me c)^2(\langle f^2_1\rangle-\langle f_1\rangle^2+\langle f^2_2\rangle-\langle f_2\rangle^2)
\label{dress2}
\end{equation}
and, consequently, the fermion effective mass in the laser field can be defined as
\begin{equation}
\bbar{m}=\sqrt{m^2+(z\mu\me)^2(\langle f^2_1\rangle-\langle f_1\rangle^2+\langle f^2_2\rangle-\langle f_2\rangle^2)}.
\label{dress3}
\end{equation}
These properties of the gauge-invariant momentum dressing can be further exploited in the finite momenta integrations that lead to the probability 
distributions of QED processes assisted by the laser fields. In particular, our analysis of resonances will become more straightforward, although 
without this modification of momentum dressing it is equally possible.

We conclude this section by noting that it is not necessary to ascribe any physical meaning to the momentum dressing (compare, e.g., 
the discussion in Refs.~\cite{HHIM2012,Reiss2014}). The latter is entirely defined by the laser-pulse averages $\langle f_j\rangle$ and 
$\langle f_1^2+f_2^2\rangle$, that in principle can be determined by means of the interferometric measurements (see, e.g., \cite{KCK2015a}). 
In the following, we shall apply a momentum dressing~\eqref{dress1a} \textit{only} as a useful mathematical tool.

\subsection{Probability amplitudes}
\label{ampli}

Consider a QED process that in the lowest order of perturbation theory is represented by a two-vertex Feynman diagram
with four external fermionic legs and an internal photon line. For now, we assume that the fermions are distinguishable. In this general situation,
the probability amplitude for the process equals
\begin{align}
\mathcal{A}(Q_1,&Q_2;Q_3,Q_4)=-4\pi\ii\alpha z_Az_B\int \dd^4 x\dd^4 y\frac{\dd^4 K}{(2\pi)^4} \nonumber \\
\times & [j_{Q_2Q_1}(x)]^\rho \ee^{-\ii K\cdot x}\frac{-g_{\rho\nu}}{K^2+\ii\epsilon}\ee^{\ii K\cdot y}[j_{Q_4Q_3}(y)]^\nu,
\label{amp1}
\end{align}
where the symbol $Q_j$ means the collection of fermionic asymptotic parameters $(\bm{p}_j,\lambda_j,\beta_j)$. Here, quantities with indices 
$j=1,3$ correspond to the incoming lines, whereas $j=2,4$ to the outgoing ones. Also, we use the shorthand notation for the Dirac current,
\begin{equation}
[j_{Q_jQ_i}(x)]^\nu = [j^{(\beta_j\beta_i)}_{\bm{p}_j\lambda_j;\bm{p}_i\lambda_i}(x)]^\nu .
\label{amp2}
\end{equation}
Additionally, $z_Ae$ and $z_Be$ are the particle charges for the fermionic lines $(1\rightarrow 2)$ and $(3\rightarrow 4)$, respectively.
From now on, we shall use the following abbreviations:
\begin{align}
[j^{(\beta_j\beta_i)}_{\bm{p}_j\lambda_j;\bm{p}_i\lambda_i;N}]^\nu=&
j_{Q_jQ_i;N}^\nu , \nonumber \\
\quad [D_{\bm{p}_j\lambda_j;\bm{p}_i\lambda_i}^{(\beta_j\beta_i)(j_1,j_2)}]^\nu=&
D_{Q_jQ_i}^{(j_1,j_2)\nu}.
\label{amp3}
\end{align}
Thus, in relation to Eq.~\eqref{fd2}, we can represent the Dirac current~\eqref{amp2} such that
\begin{equation}
[j_{Q_jQ_i}(x)]^\nu =\! \frac{m c}{V\sqrt{p^0_jp^0_i}}
\sum_{N=-\infty}^{\infty} \ee^{-\ii(\beta_i\bbar{p}_i-\beta_j\bbar{p}_j+Nk)\cdot x} j_{Q_jQ_i;N}^\nu ,
\label{amp5}
\end{equation}
where $m$ stands for the rest mass $m_A$ or $m_B$, depending on the fermionic line $(1\rightarrow 2)$ or $(3\rightarrow 4)$. In Eq.~\eqref{amp1},
we have substituted the Feynman photon propagator in the so-called Feynman gauge,
\begin{equation}
D_{\rho\nu}(x-y)=\int \frac{\dd^4 K}{(2\pi)^4}\, \ee^{-\ii K\cdot x}\frac{-g_{\rho\nu}}{K^2+\ii\epsilon}\ee^{\ii K\cdot y}.
\label{amp6}
\end{equation}
It is also worth noting that due to the conservation of the Dirac current [Eq.~\eqref{volc2}], the probability amplitude~\eqref{amp1} is gauge-invariant.
One can also see that it is determined by the Fourier transform of the Dirac current, 
\begin{equation}
\tilde{j}^{\nu}_{Q_jQ_i}(K)=\int\dd^4x\, \ee^{-\ii K\cdot x}j^{\nu}_{Q_jQ_i}(x),
\label{amp7}
\end{equation}
which for an infinite train of pulses becomes
\begin{align}
\tilde{j}^{\nu}_{Q_jQ_i}(K)=&\frac{mc}{V\sqrt{p^0_jp^0_i}}(2\pi)^4 \nonumber \\
\times&\sum_{N=-\infty}^{\infty}\delta^{(4)}(\beta_i\bbar{p}_i-\beta_j\bbar{p}_j+Nk)j^{\nu}_{Q_jQ_i;N}.
\label{amp8}
\end{align}
Here, the coefficients $j^{\nu}_{Q_jQ_i;N}$ are defined by Eqs.~\eqref{amp3}, \eqref{fd3}, and~\eqref{fd4}. This leads to the following expression 
for the probability amplitude,
\begin{align}
\mathcal{A}(Q_1,Q_2&;Q_3,Q_4)=-4\pi\ii\alpha z_Az_B(2\pi)^4\frac{m_Am_Bc^2}{V^2\sqrt{p^0_1p^0_2p^0_3p^0_4}}
\nonumber \\
\times &\sum_{N=-\infty}^{\infty}\delta^{(4)}(\beta_1\bbar{p}_1-\beta_2\bbar{p}_2+\beta_3\bbar{p}_3-\beta_4\bbar{p}_4+Nk) \nonumber \\
\times & \sum_{M=-\infty}^{\infty}j^{\rho}_{Q_2Q_1;N-M}\frac{-g_{\rho\nu}}{K^2+\ii\epsilon}j^{\nu}_{Q_4Q_3;M},
\label{amp9}
\end{align}
in which $K=\beta_3\bbar{p}_3-\beta_4\bbar{p}_4+Mk$ or $K=-(\beta_1\bbar{p}_1-\beta_2\bbar{p}_2+(N-M)k$. 

In the following, we shall assume that the probability amplitude is the sum of amplitudes corresponding to the $N$-quanta absorption or emission processes,
meaning that
\begin{equation}
\mathcal{A}(Q_1,Q_2;Q_3,Q_4)=\sum_{N=-\infty}^{\infty}\mathcal{A}_N(Q_1,Q_2;Q_3,Q_4).
\label{amp10}
\end{equation}
If we further introduce the four-vectors,
\begin{equation}
R_{ij}=\beta_i\bbar{p}_i-\beta_j\bbar{p}_j,
\label{amp11}
\end{equation}
and
\begin{equation}
\mathcal{T}_N(Q_1,Q_2;Q_3,Q_4)=\sum_{M=-\infty}^{\infty}j^{\rho}_{Q_2Q_1;N-M}\frac{-g_{\rho\nu}}{K^2+\ii\epsilon}j^{\nu}_{Q_4Q_3;M},
\label{amp12}
\end{equation}
with
\begin{equation}
K=R_{34}+Mk=-R_{12}-(N-M)k,
\label{amp13}
\end{equation}
then the probability amplitude~\eqref{amp10} can be written in a compact form,
\begin{align}
\mathcal{A}_N &(Q_1,Q_2;Q_3,Q_4)=-4\pi\ii\alpha z_Az_B\frac{m_Am_Bc^2}{V^2\sqrt{p^0_1p^0_2p^0_3p^0_4}} \nonumber \\
\times & (2\pi)^4\delta^{(4)}(R_{12}+R_{34}+Nk)\mathcal{T}_N(Q_1,Q_2;Q_3,Q_4).
\label{amp14}
\end{align}
Note that this formula is valid for QED processes that occur in the presence of an infinite train of identical pulses, in contrast to the case
of a single pulse. As will be shown below, the Ole\u{\i}nik resonances are exclusively well defined for the former. For isolated pulses, on the other hand, 
the resonances appear as finite peak-structures in probability distributions, and they are frequently accompanied by interference structures~\cite{tridentAnton2011}. 
This prevents their unambiguous identification.

\subsection{Ole\u{\i}nik resonances}
\label{generalities}

For a QED process assisted by an infinite train of laser pulses, we shall analyze below the conditions for Ole\u{\i}nik resonances.
Going back to the Fourier decomposition of the Dirac current~\eqref{fd2} and the expression describing the probability amplitude of a QED 
process~\eqref{amp1}, we conclude that the integrals over $\dd^4x$ and $\dd^4y$ result in two conservation laws,
\begin{align}
0=&\beta_1\bbar{p}_1-\beta_2\bbar{p}_2+M_1k+K, \nonumber \\
0=&\beta_3\bbar{p}_3-\beta_4\bbar{p}_4+M_2k-K,
\label{ot1}
\end{align}
with arbitrary integers $M_1$ and $M_2$. Then, the integral with respect to $\dd^4K$ in Eq.~\eqref{amp1} leads to the momentum conservation condition,
\begin{equation}
\beta_1\bbar{p}_1-\beta_2\bbar{p}_2+\beta_3\bbar{p}_3-\beta_4\bbar{p}_4+Nk=0,
\label{ot2}
\end{equation}
with $N=M_1+M_2$. Note that Ole\u{\i}nik resonances, being poles of the Feynman propagator~\eqref{amp6}, appear for such fermion momenta that $K^2=0$~\cite{Olejnik1967,OlejnikBook,OlejnikRes1,OlejnikRes2}. 
This could happen only for certain values of $M_1$ and $M_2$. Our aim is, therefore, to determine the respective kinematics. 

In our analysis we  
assume that momenta of the incoming lines (i.e., $p_1$ and $p_3$) are known (although other possibilities could be equally considered), whereas 
the remaining parameters have to be determined from the conservation conditions~\eqref{ot1}. In order to do so, we rewrite Eqs.~\eqref{ot1} such that
\begin{align}
\beta_2\bbar{p}_2=&\beta_1\bbar{p}_1+M_1k+K, \nonumber \\
\beta_4\bbar{p}_4=&\beta_3\bbar{p}_3+M_2k-K,
\label{ot3}
\end{align}
and square them. Taking into account the resonance condition $K^2=0$ and the fact that all fermion dressed momenta are on the same mass shell we arrive, 
after some algebra, at the following equation,
\begin{equation}
\hat{K}\cdot R=0.
\label{ot4}
\end{equation}
Here, $K=K^0\hat{K}$, $\hat{K}=(1,\bm{N})$ with the unit space-vector $\bm{N}$, whereas the four-vector $R$ is defined as
\begin{equation}
R=\beta_3M_2(\bbar{p}_3\cdot k)(\beta_1\bbar{p}_1+M_1k)+\beta_1M_1(\bbar{p}_1\cdot k)(\beta_3\bbar{p}_3+M_2k).
\label{ot5}
\end{equation}
Moreover, 
\begin{equation}
K^0=-\frac{\beta_1M_1\bbar{p}_1\cdot k}{\hat{K}\cdot(\beta_1\bbar{p}_1+M_1k)}=\frac{\beta_3M_2\bbar{p}_3\cdot k}{\hat{K}\cdot(\beta_3\bbar{p}_3+M_2k)}.
\label{ot6}
\end{equation}
Then, by introducing the normalized space vector $\hat{\bm{R}}=\bm{R}/|\bm{R}|$, one can rewrite Eq.~\eqref{ot4} in the form
\begin{equation}
\bm{N}\cdot\hat{\bm{R}}=\frac{R^0}{|\bm{R}|}.
\label{ot7}
\end{equation}
Hence, we conclude that Eq.~\eqref{ot4} can only be satisfied if $R$ is the space-type four-vector,
\begin{equation}
R^2=(R^0)^2-(\bm{R})^2 \leqslant 0.
\label{ot8}
\end{equation}
For the given incoming momenta $p_1$ and $p_3$, and for the given laser field parameters, this inequality determines the allowed integer
numbers $M_1$ and $M_2$. Selecting a particular pair $(M_1,M_2)$, one settles down the corresponding four-vector $R$ and proceeds to determine 
the null four-vector $K$, and momenta of the outgoing fermions $p_2$ and $p_4$ from Eq.~\eqref{ot3}.
In doing so, we choose two additional unit vectors $\hat{\bm{R}}_j$, $j=1,2$, such that together with $\hat{\bm R}$ they form 
a triad $(\hat{\bm{R}}_1,\hat{\bm{R}}_2,\hat{\bm{R}})$ of the right-handed basis, meaning that 
$\hat{\bm{R}}_1\times\hat{\bm{R}}_2=\hat{\bm{R}}$. Then, the most general solution of Eq.~\eqref{ot7} depends on an angle, 
$0\leqslant\sigma_R < 2\pi$, such that
\begin{equation}
\bm{N}=\frac{1}{|\bm{R}|}\bigl[\sqrt{-R^2}(\cos\sigma_R\hat{\bm{R}}_1+\sin\sigma_R\hat{\bm{R}}_2)+R^0\hat{\bm{R}} \bigr].
\label{ot9}
\end{equation}
Having known $\bm{N}$ and $\hat{K}$, we determine $K^0$ from Eq.~\eqref{ot6}, and the remaining dressed momenta $\bbar{p}_2$ and $\bbar{p}_4$ 
from Eq.~\eqref{ot3}; the asymptotic bare momenta $p_2$ and $p_4$ are then obtain from Eq.~\eqref{dress1b}. In order to get the one-to-one 
correspondence between the outgoing momenta and the angle $\sigma_R$ we have to uniquely define at least one of the vectors $\hat{\bm{R}}_j$. 
Specifically, we set it up such that for the given polar and azimuthal angles of $\hat{\bm{R}}$, $\theta_R$ and $\varphi_R$, we have
$\hat{\bm{R}}_1=(\cos\theta_R\cos\varphi_R,\cos\theta_R\sin\varphi_R,-\sin\theta_R)^T$. The aforementioned procedure will be used in Sec.~\ref{sec::oleinik}
to find Ole\u{\i}nik resonances in a trident process.

\begin{figure}
\includegraphics[width=6cm]{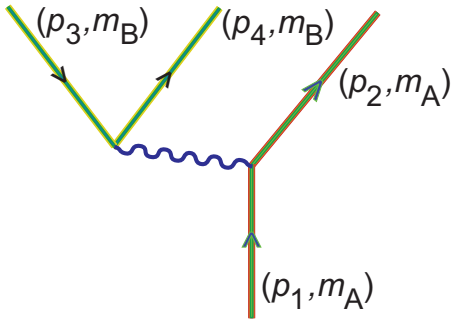}
\caption{Feynman diagram for the trident process involving distinguishable particles. All four-momenta are on the mass shells, i.e., 
$p_1\cdot p_1=p_2\cdot p_2=m_A^2c^2$ and $p_3\cdot p_3=p_4\cdot p_4=m_B^2c^2$, where $m_A$ and $m_B$ are fermionic masses.
\label{FeynmanTrident}}
\end{figure}

In closing this section we note that the fact that the probability amplitude becomes infinite at the Ole\u{\i}nik resonances originates 
from the idealized conditions that both the laser and particle beams used in this paper are described by infinite waves. It is only under such assumptions
that the position of resonances can be precisely established. In this context it is important to realize that even though experimentally
both these beams are finite in time and space, in the realm of QED it is impossible to account for those characteristics exactly. 
In our further analysis of Ole\u{\i}nik resonances, which will be performed in Sec.~\ref{trid}, 
we shall adopt therefore the prescription proposed in Ref.~\cite{HMK2010}. Namely, we shall assume that $\epsilon$ in the denominator of the photon 
propagator~\eqref{amp6} equals
\begin{equation}
\epsilon=2|K^0|/cT_0,
\label{otr1}
\end{equation}
where $T_0$ is a typical timescale of laser-matter interaction. Since the duration of a single pulse in the train is equal to $2\pi/ck^0$, we shall choose further that
\begin{equation}
\epsilon=|K^0k^0|/\pi N_0,
\label{otr2}
\end{equation}
where $N_0$ denotes the number of pulses in the train. For an infinite train of pulses, we have $T_0\rightarrow\infty$ and $N_0\rightarrow\infty$.
In actual computations, both will be kept finite but large. In closing, we stress that this prescription is only used to avoid infinities in the probability distributions
of product particles.

\section{Trident process}
\label{trid}

\subsection{Probability distributions}
\label{sec::distributions}

For the trident process represented by the Feynman diagram with two distinguished fermion lines in Fig.~\ref{FeynmanTrident}, 
we have one fermion in the remote past and three fermions in the far future. Let us assume that the incoming fermion is labeled by the multi-index $Q_1$. This means that $\beta_2=\beta_1$ (i.e., the line corresponding to the current $j_{Q_2,Q_1}$ describes the propagation of either the particle, $\beta_1=1$, or the antiparticle, $\beta_1=-1$) and $\beta_4=-\beta_3$ (i.e., the current $j_{Q_4,Q_3}$ corresponds to the particle-antiparticle creation). In our further analysis and without loosing generality we shall use interchangeably $\beta=\beta_1=\beta_2$ and $\beta=\beta_4=-\beta_3$ with $\beta=\pm$.

For fixed multi-index $Q_1$ and for very long pulse train (i.e., $T_0\rightarrow\infty$ or, equivalently, $N_0\gg 1$), 
we define the probability rate of the pair creation. For this purpose we use the standard prescription in relation to Eq.~\eqref{amp14},
\begin{align}
\bigl[(2\pi)^4\delta^{(4)}(R_{12}+ & R_{34}+Nk)\bigr]^2=cT_0V \nonumber \\ 
\times &(2\pi)^4\delta^{(4)}(R_{12}+R_{34}+Nk).
\label{tri1}
\end{align}
Since the density of final states is equal to
\begin{equation}
\sum_{\lambda_2=\pm}\frac{V\dd^3p_2}{(2\pi)^3}\sum_{\lambda_3=\pm}\frac{V\dd^3p_3}{(2\pi)^3}\sum_{\lambda_4=\pm}\frac{V\dd^3p_4}{(2\pi)^3},
\label{tri2}
\end{equation}
we obtain the total probability for the pair creation with the exchange of $N$ fundamental laser quanta of energy $\omega$,
\begin{align}
P_N=&\sum_{\lambda_2,\lambda_3,\lambda_4=\pm}\int\frac{V\dd^3p_2}{(2\pi)^3}\frac{V\dd^3p_3}{(2\pi)^3}\frac{V\dd^3p_4}{(2\pi)^3} |\mathcal{A}_N|^2 \nonumber \\
=&\frac{\alpha^2}{2\pi^3}(z_Am_Az_Bm_Bc^2)^2\frac{cT_0}{p^0_1}\sum_{\lambda_2,\lambda_3,\lambda_4=\pm}\nonumber \\
&\times  \int \frac{\dd^3p_2}{p^0_2}\frac{\dd^3p_3}{p^0_3}\frac{\dd^3p_4}{p^0_4}\delta^{(4)}(R_{12}+R_{34}+Nk) |\mathcal{T}_N|^2,
\label{tri3}
\end{align}
which allows us to define the respective rate,
\begin{equation}
W_N=\frac{P_N}{T_0}.
\label{tri4}
\end{equation}
Closely related are differential probability rates that will be defined below for a certain geometry.

We consider a target particle initially at rest ($\bm{p}_1=\bm{0}$) which is exposed to the laser pulse train propagating in the $z$-direction. 
Moreover, we assume that the polarization vectors of the laser field $\bm{\varepsilon}_1$ and $\bm{\varepsilon}_2$ are along
the $x$- and $y$-directions, respectively. Further, we fix the momentum $\bm{p}_3$ (i.e., the momentum of created positrons) 
as well as two projections of $\bm{p}_2$ onto the polarization vectors (i.e., $p^{\bot}_{2,j}=\bm{\varepsilon}_j\cdot\bm{p}_2=-\varepsilon_j\cdot p_2$, 
$j=1,2$). For such geometry, we define the six-fold probability distribution for the trident process, that depends on five continuous variables, 
$\bm{p}_3$ and $p^{\bot}_{2,j}$, and the discrete one, $N$. The latter determines the net amount of radiation energy, $N\omega$, absorbed from ($N>0$) or 
emitted to ($N<0$) the laser field. In compliance with the above, let us further assume that $Q_3$ is fixed (i.e., both $\bm{p}_3$ and $\lambda_3$, as $\beta_3$ is already fixed to be $-\beta_1$), which leads to the triply-differential rate
\begin{equation}
\frac{\dd^3W^{(1,3)}_N}{\dd^3\Gamma_3}=\frac{\alpha^2}{2\pi^3}(z_Am_Az_Bm_Bc^2)^2\frac{c}{p^0_1}\Upsilon^{(1,3)}_N,
\label{tri5}
\end{equation}
where
\begin{equation}
\dd^3\Gamma_3=\frac{\dd^3p_3}{p^0_3}
\label{tri6}
\end{equation}
is the relativistically invariant integration measure with respect to the positron momentum. Moreover,
\begin{equation}
\Upsilon^{(1,3)}_N=\sum_{\lambda_2,\lambda_4}\int\frac{\dd^3p_2}{p^0_2}\frac{\dd^3p_4}{p^0_4}\delta^{(4)}(R_{12}+R_{34}+Nk) |\mathcal{T}_N|^2,
\label{tri7}
\end{equation}
where the superscript $(1,3)$ indicates that momenta and spins of the particles 1 and 3 are fixed. Note that Eq.~\eqref{tri7} can be further simplified
due to the presence of the delta function. For this purpose, we define,
\begin{align}
u=&(\bbar{\bm{p}}_1^{\bot}-\bbar{\bm{p}}_3^{\bot})\cdot\bbar{\bm{p}}_2^{\bot}-\bbar{p}_1\cdot\bbar{p}_3+(m_Ac)^2 
\! + \! \beta Nk\cdot(\bbar{p}_1-\bbar{p}_3), \nonumber \\
w=&\frac{1}{2}\bigl[\bbar{\bm{p}}_2^{\bot}\cdot\bbar{\bm{p}}_2^{\bot}+(m_Ac)^2\bigr], \quad Q=\bbar{p}_1-\bbar{p}_3.
\label{tri10}
\end{align}
Then, we find 
\begin{equation}
\Upsilon^{(1,3)}_N=\sum_{\lambda_2,\lambda_4}\int \dd^2p_2^{\bot} \bar{\Upsilon}^{(1,3)}_N,
\label{tri11}
\end{equation}
where the explicit form of $\bar{\Upsilon}^{(1,3)}_N$ depends on solutions of the dressed four-momenta conservation condition expressed by 
the delta function in~\eqref{tri7}. Here, we meet the following options:
\begin{itemize}
\item{Option 1:}
Let us define
\begin{equation}
\Delta=u^2-4wQ^-Q^+.
\label{tri11a}
\end{equation}
If $\Delta > 0$ and $Q^-Q^+\neq 0$, then we have two solutions,
\begin{align}
S_1: &\quad  \bbar{p}_2^-=\frac{u+\sqrt{\Delta}}{2Q^+},\quad \bbar{p}_2^+=\frac{u-\sqrt{\Delta}}{2Q^-},
\nonumber \\
S_2: &\quad  \bbar{p}_2^-=\frac{u-\sqrt{\Delta}}{2Q^+},\quad \bbar{p}_2^+=\frac{u+\sqrt{\Delta}}{2Q^-},
\label{tri11b}
\end{align}
and
\begin{align}
\bar{\Upsilon}^{(1,3)}_N=\frac{1}{\sqrt{\Delta}}\bigl[&|\mathcal{T}_N(Q_1,Q_2;Q_3,Q_4)|^2\mid_{S_1}
\nonumber \\
+&|\mathcal{T}_N(Q_1,Q_2;Q_3,Q_4)|^2\mid_{S_2}\bigr].
\label{tri12}
\end{align}
Note, that $\bar{\Upsilon}^{(1,3)}_N$ becomes singular as $\Delta\rightarrow 0$. This singularity is related to the channel closing and leads to the threshold effects discussed below.

\item{Option 2:}
If $Q^+=0$ and $Q^-\neq 0$, then
\begin{equation}
S_+: \quad  \bbar{p}_2^-=\frac{wQ^-}{u},\quad \bbar{p}_2^+=\frac{u}{Q^-},
\label{tri13a}
\end{equation}
and 
\begin{equation}
\bar{\Upsilon}^{(1,3)}_N=\frac{1}{|u|}|\mathcal{T}_N(Q_1,Q_2;Q_3,Q_4)|^2\mid_{S_+}.
\label{tri13}
\end{equation}

\item{Option 3:}
If $Q^-= 0$ and $Q^+\neq 0$, then
\begin{equation}
S_-: \quad  \bbar{p}_2^-=\frac{u}{Q^+},\quad \bbar{p}_2^+=\frac{wQ^+}{u},
\label{tri14a}
\end{equation}
and 
\begin{equation}
\bar{\Upsilon}^{(1,3)}_N=\frac{1}{|u|}|\mathcal{T}_N(Q_1,Q_2;Q_3,Q_4)|^2\mid_{S_-}.
\label{tri14}
\end{equation}
\end{itemize}
At this point we note that, for the trident process, $Q^+$ and $Q^-$ cannot be simultaneously equal to zero. Also, the options 2 and 3 are met very rarely 
and, in fact, they have never occurred in our numerical analysis.

Based on the above considerations, we define the spin-resolved differential probability rate for the trident process that has been accompanied by absorption
of $N$ laser quanta from a pulse train,
\begin{align}
&\frac{\dd^5W^{(1,3)}_N(\bm{p}_1,\lambda_1;\bm{p}_3,\lambda_3;\bm{p}_2^{\bot},\lambda_2;\lambda_4)}{\dd^3\Gamma_3\dd^2p_2^{\bot}} \nonumber \\
&\qquad =\frac{\alpha^2}{2\pi^3}(z_Am_Az_Bm_Bc^2)^2\frac{c}{p^0_1}\bar{\Upsilon}^{(1,3)}_N(Q_1,Q_2;Q_3,Q_4).
\label{tri16}
\end{align}
This, in turn, allows us to define the probability rate per a single pulse from the train. For this, we multiply Eq.~\eqref{tri16}
by the pulse duration $2\pi/ck^0$ and divide it by the relative flux of initial charged particles and laser photons, 
$k\cdot p_1/k^0p_1^0$ (note that in the reference frame in which the incoming particle is at rest this extra factor is 1). Hence, we obtain
\begin{align}
&\frac{\dd^5P^{(1,3)}_N(\bm{p}_1,\lambda_1;\bm{p}_3,\lambda_3;\bm{p}_2^{\bot},\lambda_2;\lambda_4)}{\dd^3\Gamma_3\dd^2p_2^{\bot}} \nonumber \\
&\qquad =\frac{\alpha^2}{\pi^2}\frac{(z_Am_Az_Bm_Bc^2)^2}{k\cdot p_1}\bar{\Upsilon}^{(1,3)}_N(Q_1,Q_2;Q_3,Q_4),
\label{tri16a}
\end{align}
In our further analysis we shall not investigate the spin effects. Therefore, the above distribution is summed over the final particle spin degrees of freedom 
and averaged over the initial one, which leads to 
\begin{align}
&\frac{\dd^5P^{(1,3)}_N(\bm{p}_1;\bm{p}_3;\bm{p}_2^{\bot})}{\dd^3\Gamma_3\dd^2p_2^{\bot}} \nonumber \\
&\quad =\frac{1}{2}\sum_{\lambda_1,\lambda_2,\lambda_3,\lambda_4=\pm}\frac{\dd^5P^{(1,3)}_N(\bm{p}_1,\lambda_1;\bm{p}_3,\lambda_3;\bm{p}_2^{\bot},\lambda_2;\lambda_4)}{\dd^3\Gamma_3\dd^2p_2^{\bot}}.
\label{tri16b}
\end{align}
Note, that these are the relativistically invariant distributions as the perpendicular components of $p_2$ are always defined with respect to the polarization vectors of the laser beam, i.e., $(\bm{p}^{\bot}_2)_j=-\varepsilon_j\cdot p_2$ for $j=1,2$.

\begin{figure}
\includegraphics[width=8cm]{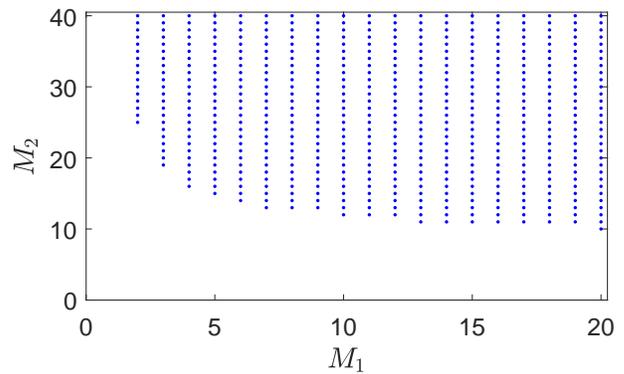}
\caption{Pairs of integers $(M_1,M_2)$ for which Ole\u{\i}nik resonances appear for $\sigma_R=0$, $\bm{p}_1=\bm{0}$ (in the muon rest frame), 
positron kinetic energy $E_3=cp_3^0-m_{\mathrm{e}}c^2=m_{\mathrm{e}}c^2$, and its polar and azimuthal 
angles, $\theta_3=0.3\pi$ and $\varphi_3=\pi$. Details of the laser pulse parameters are discussed in the text. Dots indicate the existence of 
resonances for $-20\leqslant M_1\leqslant 20$ and $-40\leqslant M_2\leqslant 40$. The asymmetric distribution is mainly due to the different 
masses of muons and electrons.
\label{resdiscr}}
\end{figure}
\begin{figure}
\includegraphics[width=8cm]{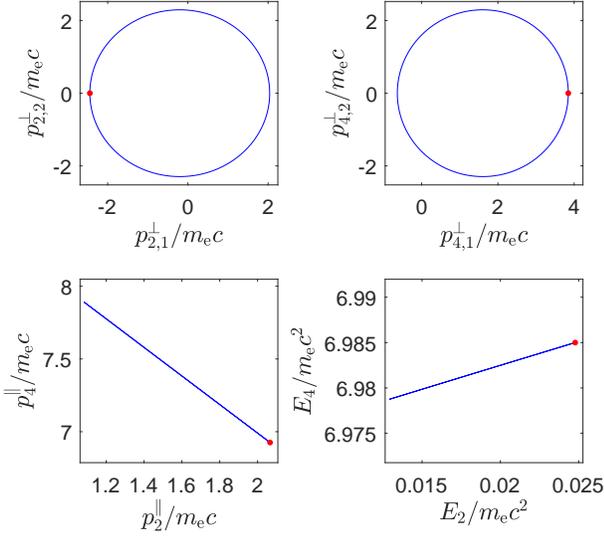}
\caption{Some particular projections of the resonance curve for $(M_1,M_2)=(5,15)$ from Fig.~\ref{resdiscr} embedded in the six-dimensional 
space $(\bm{p}_2,\bm{p}_4)$ (with fixed $\bm{p}_3$). The red dot indicate momenta for the angle $\sigma_R=0$ and, with increasing $\sigma_R$,
it will move counterclockwise in the top panels. In the bottom right panel correlations between kinetic energies of the final muon $E_2$ and created 
electron $E_4$ are also presented, showing that even though components of momenta can change significantly, nevertheless, their kinetic energies remain nearly 
constant. Note that the muon kinetic energy is much smaller than the electron kinetic energy, as the muon is a much heavier particle.
\label{ressigma}}
\end{figure}

In closing this section, we remark that the above distributions are suitable for the case of distinguishable particles with different masses $m_A$ and $m_B$,
or different charges $z_Ae$ and $z_Be$. For indistinguishable particles, we have to put $m_A=m_B$, $z_A=z_B$ and antisymmetrize amplitudes with respect 
to momenta $p_2$ and $p_4$. In this case, the amplitude $\mathcal{T}_N(Q_1,Q_2;Q_3,Q_4)$ that appears in Eq.~\eqref{tri3} (as well as in the proceeding
formulas) should be replaced by
\begin{equation}
\frac{1}{\sqrt{2}}\bigl[\mathcal{T}_N(Q_1,Q_2;Q_3,Q_4)-\mathcal{T}_N(Q_1,Q_4;Q_3,Q_2)\bigr].
\label{tri17}
\end{equation}
This relates, for instance, to the situation when $e^-e^+$ pairs are created in laser-field--electron collisions, as realized experimentally in SLAC~\cite{Stanford1,Stanford2}.
In this paper, however, we consider a different incoming particle so there is no necessity to  
anti-symmetrize the probability amplitude of the process (as it was done, for instance, for the M{\o}ller scattering in Refs.~\cite{PanekMoller2004,PanekMoller2004a,AtomsFelipe2019}). 
As the example we consider the muon of mass $m_{\mu}\approx 206.768m_{\mathrm{e}}$. Note that fundamental QED processes with muons have been already 
studied in literature~\cite{muon1,muon2,muon3,muon4,muon5}, but not in the context of Ole\u{\i}nik resonances and threshold effects.

\subsection{Ole\u{\i}nik resonances}
\label{sec::oleinik}

As follows from our analysis in Sec.~\ref{generalities}, the positions of resonances depend on the momenta $p_1$ and $p_3$ incoming to the Feynmam 
diagram. Since $p_1$ represents the colliding muon at rest, therefore, the positron four-momentum $p_3$ has to be settled. By doing this and 
by fixing an angle $\sigma_R$ we can determine all integer pairs $(M_1,M_2)$ leading to resonances. For particular choices of $p_3$ and $\sigma_R$ 
the results are presented in Fig.~\ref{resdiscr}. The electric field of the laser pulse is defined by the master function~\eqref{lp10} with the 
normalization constant $\mathcal{N}=1$. The remaining parameters are: $\mu=10$, $N_{\mathrm{osc}}=2$, $\omega_L=N_{\textrm{osc}}\omega=m_{\mathrm{e}}c^2$, 
and $\chi=0$. With these parameters we find that $\langle f_1\rangle \approx 0.083$ and $\langle f^2_1\rangle \approx 0.073$. Since dots 
in Fig.~\ref{resdiscr} fill almost the entire first quarter of the plane $(M_1,M_2)$, resonances should quite frequently appear in the probability 
distribution for the trident process.

Each resonance indicated by a dot in Fig.~\ref{resdiscr} can be represented in the six-dimensional space of momenta $(\bm{p}_2,\bm{p}_4)$ 
(with fixed $\bm{p}_3$) by a curve that is parametrized by the angle $\sigma_R$. In fact, if we allow  the positron momentum $\bm{p}_3$ to change as well
then such resonance will be described by a four-dimensional manifold embedded in the nine-dimensional space of final momenta, making their 
analysis very cumbersome. For this reason, we rather fix the positron momentum ${\bm p}_3$. In that case,  
we present in Fig.~\ref{ressigma} projections of the resonance curve for $(M_1,M_2)=(5,15)$ on some particular planes in the six-dimensional $(\bm{p}_2,\bm{p}_4)$ 
space, as well as the kinetic energy correlation 
for particles 2 and 4 (i.e., final muon and created electron, respectively). The latter shows that, for this resonance, 
the kinetic energies of final particles change within small intervals (are nearly constant) although the components of their momenta can change  
significantly, i.e, even by few $m_{\mathrm{e}}c$.

\begin{figure}
\includegraphics[width=7cm]{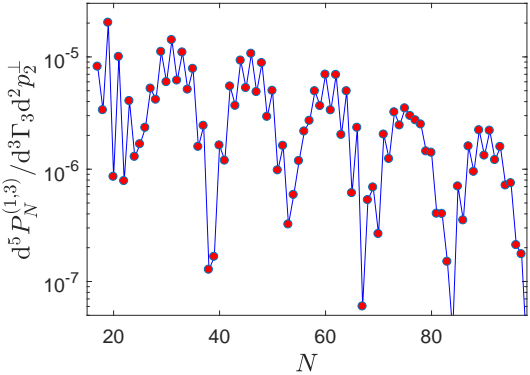}
\caption{Spin-averaged probability distribution of the trident process~\eqref{tri16b} as a function of the number of laser quanta $N$ 
absorbed from the laser field for $\bm{p}_1=\bm{0}$, $E_3=cp_3^0-m_{\mathrm{e}}c^2=m_{\mathrm{e}}c^2$, $\theta_3=0.3\pi$, $\varphi_3=\pi$, 
and $\bm{p}_2^{\bot}=\bm{0}$. The laser field parameters are: $\mu=10$, $N_{\mathrm{osc}}=2$, $\omega_L=N_{\textrm{osc}}\omega=m_{\mathrm{e}}c^2$, 
and $\chi=0$. The solid line connecting the dots is plotted to guide an eye.
\label{figN}}
\end{figure}

\subsection{Resonances in energy distributions}
\label{sec::resonances}

In this Section, we consider the trident process in which a muon collides with an infinite train of pulses. We choose the reference frame where the muon 
is initially at rest (${\bm p}_1={\bm 0}$) and the linearly polarized light [with $\bm{\varepsilon}_1=\bm{e}_x$ and $f_2(k\cdot x)=0$] propagates in the $z$-direction. 
Pulses comprising the train are described by Eq.~\eqref{lp10}, where we further assume that $\mu=10$, $N_{\mathrm{osc}}=3$, $\omega_L=N_{\mathrm{osc}}\omega=m_{\mathrm{e}}c^2$, 
$\chi=\pi/2$, and $\mathcal{N}=1$. This means that the maximum of the electric field is $\mu/N_{\mathrm{osc}}=10/3$ of the Sauter-Schwinger electric field strength unit, $\mathcal{E}_S=m^2_{\mathrm{e}}c^3/(|e|\hbar)$. 
In addition, we keep the final muon transverse momentum equal to zero (${\bm p}_2^\perp={\bm 0}$) and we fix the positron momentum ${\bm p}_3$ such that 
$E_3=cp_3^0-m_{\mathrm{e}}c^2=m_{\mathrm{e}}c^2$, $\theta_3=0.3\pi$, and $\varphi_3=\pi$. For such conditions, we plot in Fig.~\ref{figN} the dependence 
of the probability distribution~\eqref{tri16b} on the number of laser photons $N$ absorbed from the field. We observe the modulation of the distribution 
with the period around $\Delta N=15$. If we decrease the kinetic energy $E_3$, the period of those oscillations decreases as well. Finally, at 
$E_3=0.3m_{\mathrm{e}}c^2$ we observe (up to small changes on the logarithmic scale) a monotonic decrease of probability distribution with $N$. On the other hand, 
for larger $E_3$ the period $\Delta N$ also increases. Such a behavior shows that the modulation present in Fig.~\ref{figN} is not related to the 
multiphoton absorption, neither to resonances (the pattern marginally depends on the change of $N_0$, which in all figures is set to 200), but rather 
to interference of probability amplitudes. Note that such interference modulations are typical for the Compton~\cite{Compton1,Compton2,Compton3,Compton4,Compton5,Compton6,Compton7} or Breit-Wheeler \cite{KK2012b,BW1,BW2,BW3,BW4} processes as well.

\begin{figure}
\includegraphics[width=7cm]{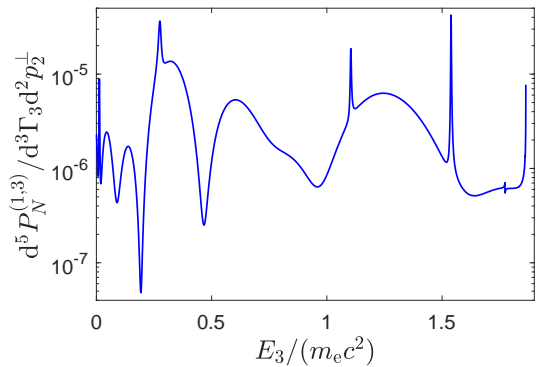}
\caption{Dependence of the probability distribution~\eqref{tri16b} on the positron kinetic energy, $E_3=cp_3^0-m_{\mathrm{e}}c^2$, for $N=20$.
The remaining parameters are the same as in Fig.~\ref{figN}. Spikes in the spectrum indicate the presence of resonances.
\label{figE}}
\end{figure}

\begin{figure}
\includegraphics[width=8cm]{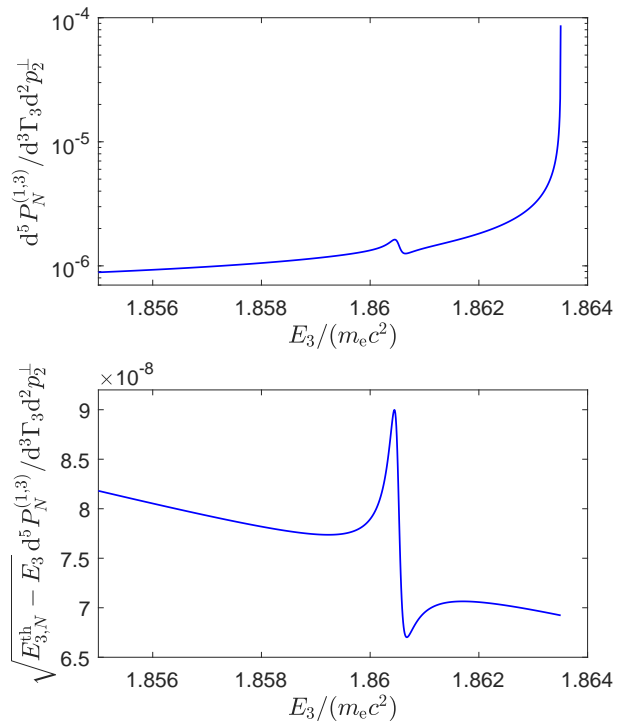}
\caption{The same as in Fig.~\ref{figE}, but with the positron kinetic energies $E_3$ closer to the threshold 
$E_{3,N}^{\mathrm{th}}\approx 1.8635\, m_{\mathrm{e}}c^2$ for the channel closing with absorption of $N=20$ fundamental laser quanta. 
In the upper panel we present the probability distribution~\eqref{tri16b} in the logarithmic scale, showing the singular behavior close to the threshold energy and the tiny Ole\u{\i}nik resonance for energies just above $1.86m_{\mathrm{e}}c^2$. In the lower panel the probability distribution is multiplied by $\sqrt{E_{3,N}^{\mathrm{th}}-E_3}$, which eliminates the singularity.
\label{figEth2}}
\end{figure}

In Fig.~\ref{figE}, we plot the energy distribution of positrons as a function of their kinetic energy $E_3$ for the fixed value of $N=20$
(with the remaining parameters kept the same as in Fig.~\ref{figN}). The latter denotes a given channel for pair creation which is open for as long as 
$N\hbar\omega\geqslant 2\bbar{m}_{\rm e}c^2$, where $\bbar{m}_{\rm e}$ is given by Eq.~\eqref{dress3}. This inequality explains why in Fig.~\ref{figE}
the distribution abruptly vanishes. Namely, $N$ fundamental laser quanta of energy $\omega$ becomes insufficient to create pairs with energies 
larger than the threshold energy for that channel, $E_{3,20}^{\rm th}$. Mathematically, such channel closing is due to vanishing of the discriminant 
$\Delta$ defined by Eq.~\eqref{tri11a} or possibly $u$ in Eq.~\eqref{tri10} for the remaining two options, which however have never appeared in our numerical explorations. Since the probability distribution contains the square root of $\Delta$ in the denominator, therefore, close to the threshold energy this distribution shows up the singularity exhibited in Fig.~\ref{figE} and enhanced (by making the grid of calculations smaller) in Fig.~\ref{figEth2} in the upper panel. However, if multiplied by $\sqrt{E_{3,N}^{\mathrm{th}}-E_3}$ the distribution becomes finite (see, the bottom panel), which proves that the threshold singularity is integrable.

\subsection{Carrier-envelope-phase effects}
\label{sec::CEP}

Interaction of matter with a laser field crucially depends on the space and time properties of the latter. This is mostly due to quantum interference 
which commonly occurs in light-induced and light-assisted processes. It is important to realize that interference effects can, in principle, be 
significantly enhanced or suppressed by various factors such as the relative phases of multichromatic field components or the carrier envelope phase
of the laser pulse (see, e.g.,~\cite{Ehlotzky2001}). This offers an opportunity of coherent phase control of quantum processes, which was originally applied 
in molecular physics as means to manipulate chemical reactions (see, e.g.,~\cite{Shapiro2003}). In recent years, phase effects specific to relativistic 
regime of laser-matter interactions have been also thoroughly studied. This includes particle scattering~\cite{cep4,cep7,cep3}, the Kapitza-Dirac 
effect~\cite{cep5,cep6}, the Compton and Thomson scattering~\cite{cepp1,cepp2,cepp3}, and nonlinear pair production~\cite{cep1,cep2,BW5}. Interestingly, it has
been also demonstrated that a high sensitivity of those processes to the laser field can serve as a measure of 
field properties~\cite{cepp1,cepp3,Deeksha}.

\begin{figure}
\includegraphics[width=8.4cm]{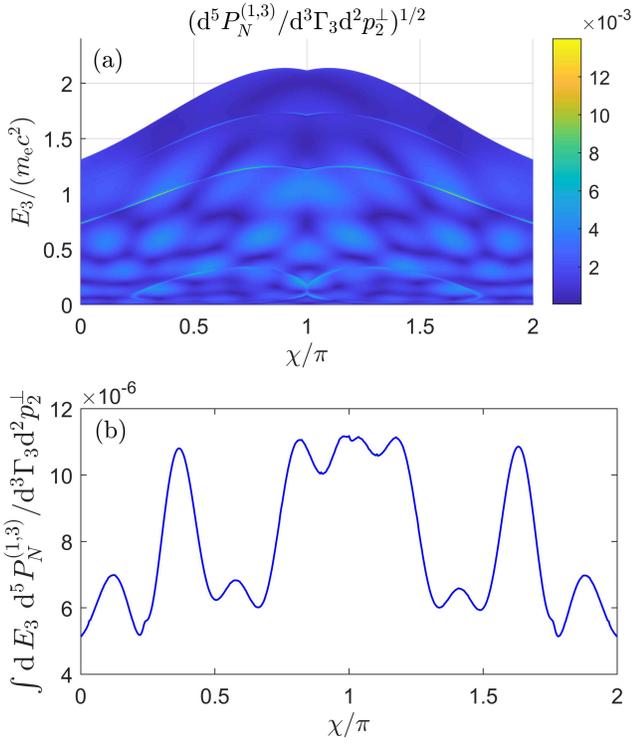}
\caption{In the upper panel, we present the two-dimensional probability distribution~\eqref{tri16b} for $N=20$ (raised to the power 1/2 for visual purposes)
as a function of the positron kinetic energy $E_3$ and the laser pulse CEP. The remaining parameters are the same as in Fig.~\ref{figN}. 
In the bottom panel, the probability distribution integrated over $E_3$ shows a significant dependence on the carrier envelope phase $\chi$.
\label{figcep}}
\end{figure}

In Fig.~\ref{figcep}, we present the dependence of the color mapping of the probability distribution as a function of the kinetic energy 
of created positron, $E_3$, and the carrier 
envelope phase of an individual pulse from the train, $\chi$. The laser field configuration is the same as before, meaning that the averages 
$\langle f_1\rangle$ and $\langle f^2_1\rangle$ that determine momentum dressing and properties of resonances do not vanish and depend on the 
phase $\chi$. For this reason, despite the smoothly varying background distribution, we observe sharp resonant peaks (the lighter lines) and 
threshold singularities (present at the border, but not sufficiently displayed with this resolution), 
the position and strength of which are, in general, $\chi$-dependent. In addition, if integrated over the positron kinetic energy $E_3$, the distribution 
exhibit a strong dependence on the carrier envelope phase. This shows that the CEP plays a significant role in the trident process.

\subsection{Lorentz meets Fano}
\label{sec::Lorentz}

Resonance phenomena are among the most fascinating and important in physics~\cite{res0,res1,res2}. The resonance scattering of elementary particles 
with matter, both in the absence and presence of the radiation background, provides the fundamental information about the properties and structure 
of elementary particles themselves, and also of solids, atoms or nuclei. Note that the interference of the background and the resonant contributions
to the probability amplitudes lead very frequently to a complicated dependence of probabilities, or cross sections, on the final particles momenta. As functions of energy, 
these probabilities very rarely exhibit the pure Lorentz-Breit-Wigner shapes, which makes it difficult to determine the resonance width and position. 
On the other hand, laser pulses depend, in principle, on various parameters which can be used to control physical phenomena. 
As mentioned above, such control can be accomplished with the help of CEP or relative phases
of multichromatic waves. Such investigations cover various topics -- from the atomic and 
solid state physics~\cite{cep4,cep7} to the relativistic strong field QED~\cite{cepp1,cepp2,cepp3,cep1,cep2,cep3,cep5,cep6,cep8}; the latter being 
the subject of our paper. Interestingly, it was predicted theoretically~\cite{LorentzFano1} that the laser phases can be used for 
filtering resonance processes in such a way that for particular laser field configurations the pure Lorentz-Breit-Wigner shape of the scattering 
resonance is restored. It has been also shown experimentally and analyzed theoretically in Ref.~\cite{LorentzFano2} that a similar situation can be observed in 
photoexcitation, in regard to which the term `\textit{Lorentz meets Fano}' has been coined. The aim of this section is to investigate a possibility 
for Ole\u{\i}nik resonances to occur in the trident process.

A general form of the multichannel scattering matrix in the close vicinity of a resonance is derived by assuming that resonances correspond 
to isolated poles of the scattering matrix. Realizing that the electromagnetic interactions are invariant under time reversal, one obtains the following 
parametrization (also called the Breit-Wigner formula) of the scattering matrix from the channel `$\mathrm{i}$' to the channel `$\mathrm{f}$' \cite{res0,res2,res3},
\begin{equation}
T_{\mathrm{fi}}=B_{\mathrm{fi}}+\frac{1}{2\pi}\frac{\sqrt{\Gamma_{\mathrm{f}}\Gamma_{\mathrm{i}}}}{E-E_R+\ii\Gamma/2}\ee^{\ii(\phi_{\mathrm{f}}+\phi_{\mathrm{i}})},
\label{ress1}
\end{equation}
where $T_{\mathrm{fi}}$ is called the $T$-matrix~\cite{res2}. It is assumed that $B_{\mathrm{fi}}$ describes the scattering background that
 marginally depends on the initial and final momenta in the vicinity of the resonance. The phases $\phi_{\mathrm{f}}$ and $\phi_{\mathrm{i}}$ 
 depend in general on both the scattering background and the structure of the resonance. Additionally, $\Gamma$ and $E_R$ are the width and 
 position of the resonance, whereas $\Gamma_{\ell}$ is the partial width for the channel $\ell$. Note that 
\begin{equation}
\sum_{\ell}\Gamma_{\ell}=\Gamma,
\label{ress2}
\end{equation}
where the summation runs over all open channels. The general formula~\eqref{ress1} can be also presented in a more compact form~\cite{res4},
\begin{equation}
T_{\mathrm{fi}}=B_{\mathrm{fi}}\frac{E-E_R+Q_{\rm fi}}{E-E_R+\ii\Gamma/2},
\label{ress2a}
\end{equation}
with a complex $Q_{\rm fi}$. For some values of parameters entering Eq.~\eqref{ress1}, the quantity $Q_{\rm fi}$ becomes real. In that case, we obtain 
the Fano formula for the cross 
section~\cite{LorentzFano2},
\begin{equation}
\sigma=\sigma_0\frac{(\varepsilon+q)^2}{\varepsilon^2+1}, \quad{\rm where}\quad \varepsilon=\frac{E-E_R}{\Gamma/2},
\label{ress3}
\end{equation}
with real $q$. The resonance in the lower panel of Fig.~\ref{figEth2} shows approximately a Fano-type shape~\cite{Fano1,Fano2}, but not with 
real $Q_{\rm fi}$ in Eq.~\eqref{ress2a} as the probability distribution does not vanish in the vicinity of the resonance energy, i.e., for $\varepsilon=-q$.

\begin{figure}
\includegraphics[width=8cm]{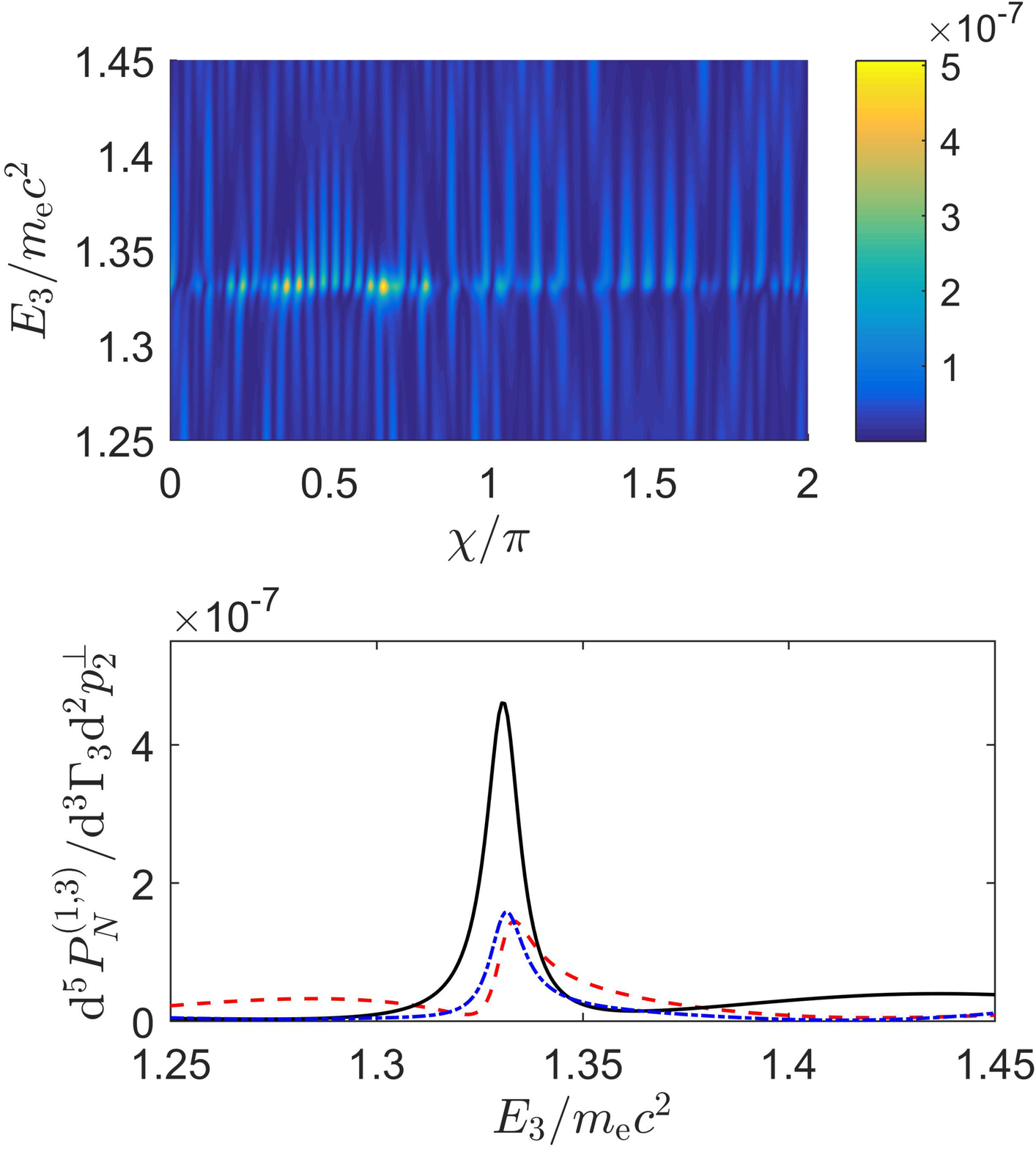}
\caption{Color mapping of the probability distribution~\eqref{tri16b} for $N=80$ (upper panel) as a function of the carrier envelope phase $\chi$ 
and the kinetic energy of created positrons $E_3=cp^0_3-m_{\mathrm{e}}c^2$. We assume that the polar and azimuthal positron ejection angles are 
$\theta_3=\pi/2$ and $\varphi_3=\pi$, respectively, and $\bm{p}^{\bot}_2=\bm{0}$. The incident fermion is the muon with rest mass 
$m_A=m_{\mu}\approx 206.768m_{\mathrm{e}}$. The laser pulse parameters are the following: $\mu=10$, $\omega_L=m_{\mathrm{e}}c^2$, $N_{\mathrm{osc}}=3$, 
with the vector potential shape function normalized such that $\langle f_1^2\rangle=\frac{1}{2}$. The latter guarantees that the position of the resonance 
on the energy scale is $\chi$-independent. In the lower panel, we present this distribution for selected CEP values: $\chi=0.6\pi$ (dashed line), 
$0.676\pi$ (solid line), and $0.76\pi$ (dot-dashed line).
\label{fll2p1res}}
\end{figure}

For the trident process considered here, all unknown \textit{a priori} parameters in~\eqref{ress1} depend on the laser field, particularly on its carrier 
envelope phase $\chi$. It might happen that, for selected values of $\chi$, the interference between the background term
$B_{\mathrm{fi}}$ and the resonant term vanishes (or becomes negligibly small), leading to the Lorentz-Breit-Wigner shape,
\begin{equation}
|T_{\mathrm{fi}}|^2\approx |B_{\mathrm{fi}}|^2+\frac{1}{4\pi^2}\frac{\Gamma_{\mathrm{f}}\Gamma_{\mathrm{i}}}{(E-E_R)^2+\Gamma^2/4}.
\label{ress4}
\end{equation}
It would be very difficult, or even impossible, to analytically determine such a phase for a given laser pulse shape. 
In order to do that, we shall proceed with numerical analysis. 

From now on, we settle the laser field parameters such that the resonance energy is independent of the carrier envelope phase. 
This can be achieved if the function~\eqref{lp10} defines the vector potential~\eqref{lp1}, i.e., $f_1(\phi)=F(\phi)$ 
whereas $f_2(\phi)=0$. Moreover, the normalization constant $\mathcal{N}$ in Eq.~\eqref{lp10} is chosen such that $\langle f_1^2\rangle=\frac{1}{2}$
for all $\chi$. For the laser pulse described above, in Fig.~\ref{fll2p1res} we present the probability distribution for the trident process in the vicinity 
of Ole\u{\i}nik resonance that occurs for energy around $1.33\, m_{\mathrm{e}}c^2$. The color mapping in the upper panel shows the probability
distribution as a function of the positron energy $E_3$ and the carrier envelope phase $\chi$. We observe here a typical interference structure 
which, for certain values of $\chi$, is either resonantly enhanced (constructive interference) or suppressed 
(destructive interference). In the lower panel of Fig.~\ref{fll2p1res}, details of such behavior for the given values of $\chi$ are shown.
We observe that relatively small changes of CEP significantly modify the resonant structure. Specifically, as suggested above, it
follows a nearly pure Lorentz-Breit-Wigner shape from which one can estimate the resonance position and width. Note also that, as expected, 
while avoiding the resonance singularity, its width depends on the duration of the laser pulse train and depends on $N_0$.

\begin{figure}
\includegraphics[width=8cm]{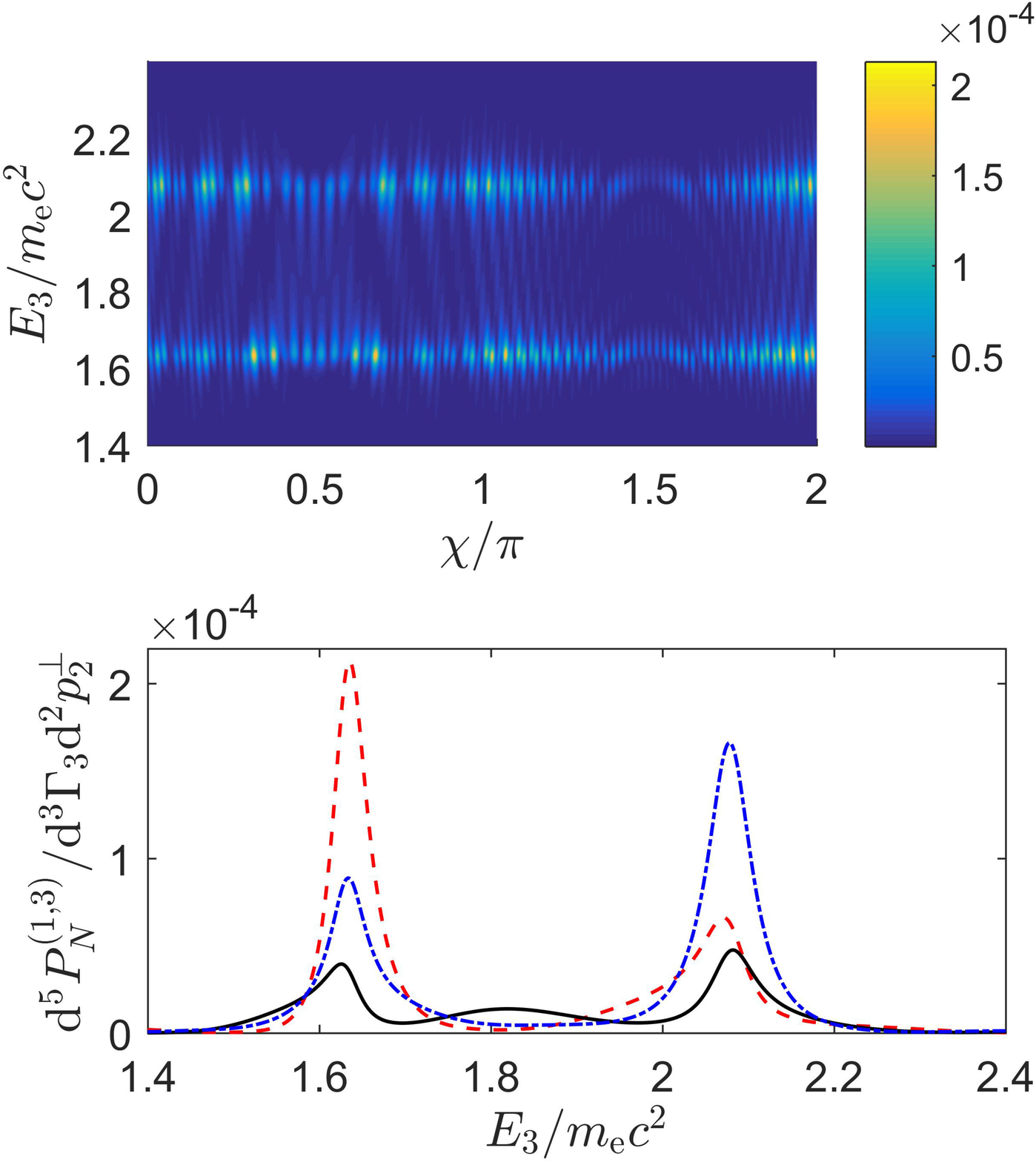}
\caption{The same as in Fig.~\ref{fll2p1res} except that now $\theta_3=0.3\pi$ and $\varphi_3=\pi$, whereas $N_{\mathrm{osc}}=2$. 
In addition, in the lower panel the selected CEPs are: $\chi=1.932\pi$ (dashed line), $1.952\pi$ (solid line), and $1.98\pi$ (dot-dashed line).
\label{fll2p2res}}
\end{figure}

By changing the laser field parameters we can also investigate closely separated resonances, as shown in Fig~\ref{fll2p2res}. 
This time we observe two resonances of energies around $1.65\, m_{\mathrm{e}}c^2$ and $2.07\, m_{\mathrm{e}}c^2$, i.e., nearly separated by  
the fundamental laser quanta of energy $\omega=\omega_L/N_{\mathrm{osc}}=0.5\, m_{\mathrm{e}}c^2$. As before, the pattern exhibits strong constructive 
and destructive interference effects. Nevertheless, it is possible to resolve the resonant structures. The point however is that while one resonance 
is purified the second one decays due to destructive interference. Thus, we conclude that high precision is required for the CEP control
of resonant structures in the trident process.

\section{Conclusions}
\label{conclusions}

We have provided a very general theoretical framework for strong-field QED processes that, in the first order of perturbation theory,
can be represented by a two-vertex Feynman diagram with four external fermion legs and a single photon line. Assuming that the processes occur
in the presence of an arbitrary train of pulses, we have developed formulas for the probability amplitude and the resulting probability distributions. 
A scheme of investigating Ole\u{\i}nik resonances, arising from the poles of the Feynman photon propagator, has been also developed. While our 
theory is applicable to a variety of processes, including pair production and annihilation, 
M\"oller scattering, Bhahba scattering, etc., we have illustrated it for a trident process.

We have considered a trident process of $e^-e^+$ pair creation that occurs in the muon--laser-field collisions. For the laser field parameters
chosen in the paper, we have observed a variety of resonances in the energy spectrum of created positrons. We have seen that their analysis can be
obscured by quantum interference effects. The latter can be controlled, however, by the carrier envelope phase of pulses in a train.
We have demonstrated that by changing the CEP we can transform the probability distribution shape into the Lorentz-Breit-Wigner one. 
This, in turn, allows one to determine the resonance position and width. Thus, our analysis allows one to unambiguously 
localize resonances in the signal of created particles in a train of pulses. Hence, it can serve as an important initial step 
toward analysis of resonances in the trident process that occurs in an isolated laser pulse. While such systematic study goes beyond the scope 
of this paper, it will be presented elsewhere.

We have also investigated the threshold behavior of the probability distributions
in the trident process. Specifically, we have shown that near the multiphoton thresholds the probability
distribution of pair creation exhibits an integrable singularity.

In closing, we stress that the presented methods and results can be also generalized to other types of two-vertex QED processes
that occur in the presence of a laser field. Specifically, to those described by the Feynman diagram with two fermion legs and two photon legs, 
including the one corresponding to the interaction with an external potential. As examples, one can mention for instance the laser-assisted Bethe-Heitler process
of pair creation~\cite{BHRos} and bremshtrahlung~\cite{bremRos}.

\section*{Acknowledgements}

This work has been supported by the National Science Centre (Poland) under Grant Nos. 2014/15/B/ST2/02203 and 2018/31/B/ST2/01251.


\begin{thebibliography}{99}

\bibitem{MTB2006}
G. A. Mourou, T. Tajima, and S. V. Bulanov, Rev. Mod. Phys. \textbf{78}, 309 (2006).

\bibitem{EKK2009}
F. Ehlotzky, K. Krajewska, and J. Z. Kami\'nski, Rep. Prog. Phys. \textbf{72}, 046401 (2009).

\bibitem{PMHK2012}
A. Di~Piazza, C. M\"uller, K. Z. Hatsagortsyan, and C.~H.~Keitel, Rev. Mod. Phys. \textbf{84}, 1177 (2012).

\bibitem{TitovRev}
A. I. Titov, B. K\"ampfer, A. Hosaka, and H. Takabe, Phys. Part. Nucl. \textbf{47}, 456 (2016).

\bibitem{HartinRev}
A. Hartin, Int. Jour. Mod. Phys. A \textbf{33} 1830011 (2018).

\bibitem{Gonoskov}
A. Gonoskov, T. G. Blackburn, M. Marklund, and S. S. Bulanov, Rev. Mod. Phys. \textbf{94}, 045001 (2022). 

\bibitem{Fedotovnew}
A. Fedotov, A. Ilderton, F. Karbstein, B. King, D. Seipt, H. Taya, and G. Torgrimsson, arXiv:2203.00091v1.

\bibitem{Stanford1}
D. L. Burke, R. C. Field, G. Horton-Smith, J. E. Spencer, D. Walz, S. C. Berridge, W. M. Bugg, K. Shmakov, A. W. Weidemann,
C. Bula, K. T. McDonald, E. J. Prebys, C. Bamber, S. J. Boege, T. Koffas, T. Kotseroglou, A. C. Melissinos, D. D. Meyerhofer,
D. A. Reis, W. Ragg, Phys. Rev. Lett. \textbf{79}, 1626 (1997).

\bibitem{Stanford2}
C. Bamber, S. J Boege, T. Koffas, T. Kotseroglou, A. C. Melissinos, D. D. Meyerhofer, D. A. Reis, W. Ragg, C. Bula, K. T. McDonald,
E. J. Prebys, D. L. Burke, R. C. Field, G. Horton-Smith, J. E. Spencer, D. Walz, S. C. Berridge, W. M. Bugg, K. Shmakov, A. W. Weidemann,
Phys. Rev. D \textbf{60}, 092004 (1999).

\bibitem{Luxe1}
https://luxe.desy.de/

\bibitem{Luxe2}
H. Abramowicz, U. Acosta, M. Altarelli, {\it et al.}, Eur. Phys. J. Spec. Top. {\bf 230}, 2445 (2021).

\bibitem{Facet1}
https://facet-ii.slac.stanford.edu/

\bibitem{Facet2}
V. Yakimenko, L. Alsberg, E. Bong, G. Bouchard, C. Clarke, C. Emma, S. Green, C. Hast, M. J. Hogan,
J. Seabury, N. Lipkowitz, B. OShea, D. Storey, G. White, G. Yocky, Phys. Rev. Accel. Beams {\bf 22}, 101301 (2019).



\bibitem{HMK2010}
H. Hu, C. M\"uller, and C. H. Keitel, Phys. Rev. Lett. \textbf{105}, 080401 (2010).

\bibitem{tridentKK2010}
K. Krajewska and J. Z. Kami\'nski, Phys. Rev. A \textbf{82}, 013420 (2010).

\bibitem{kkraj2011}
K. Krajewska, Laser Phys. \textbf{21}, 1275 (2011).

\bibitem{tridentAnton2011}
A. Ilderton, Phys. Rev. Lett. \textbf{106}, 020404 (2011).

\bibitem{King1}
B. King and H. Ruhl, Phys. Rev. D {\bf 88}, 013005 (2013).

\bibitem{tridentPRA89}
H. Hu and J. Huang, Phys. Rev. A \textbf{89}, 033411 (2014).

\bibitem{tridentKK2015}
K. Krajewska and J. Z. Kami\'nski, J. Phys.: Conf. Ser. \textbf{594}, 012024 (2015).

\bibitem{Dinu1}
V. Dinu and G. Torgrimsson, Phys. Rev. D {\bf 97}, 036021 (2018).

\bibitem{King2}
B. King and A. M. Fedotov, Phys. Rev. D {\bf 98}, 016005 (2018). 

\bibitem{Mackenroth}
F. Mackenroth and A. Di Piazza, Phys. Rev. D {\bf 98}, 116002 (2018).

\bibitem{Acosta2019}
U. H. Acosta and B K\"ampfer, Plasma Phys. Control. Fusion \textbf{61} 084011 (2019).

\bibitem{Dinu2}
V. Dinu and G. Torgrimsson, Phys. Rev. D {\bf 101}, 056017 (2020).

\bibitem{Dinu3}
V. Dinu and G. Torgrimsson, Phys. Rev. D {\bf 102}, 016018 (2020).

\bibitem{Torg1}
G. Torgrimsson, Phys. Rev. D {\bf 102}, 096008 (2020).

\bibitem{Torg2}
G. Torgrimsson, Phys. Rev. D {\bf 102}, 116008 (2020).



\bibitem{Olejnik1967}
V. P. Ole\u{\i}nik, Zh. \`Eksp. Teor. Fiz. \textbf{52}, 1049 (1967) [Sov. Phys.
JETP \textbf{25}, 697 (1967)].

\bibitem{OlejnikBook}
V. P. Ole\u{\i}nik and I. V. Belousov, \textit{Problems of the Quantum Electrodynamics of the Vacuum, Dispersive Media and Strong Fields}, (Shtintsa, Kishiniev, 1983) (in Russion).

\bibitem{OlejnikRes1}
J. B\"os, W. Brock, H. Mitter, and Th. Schott, J. Phys. A \textbf{12}, 715 (1979).

\bibitem{OlejnikRes2}
J. B\"os, W. Brock, H. Mitter, and Th. Schott, J. Phys. A \textbf{12}, 2573 (1979).

\bibitem{rosh1}
S. P. Roshchupkin, Laser Physics \textbf{6}, 837 (1996).

\bibitem{rosh2}
S. P. Roshchupkin, E. A. Padusenko, and A. I. Voroshilo, Laser Physics \textbf{22}, 1113 (2012).

\bibitem{rosh3}
A. A. Lebed' and S. P. Roshchupkin, Phys. Rev. A \textbf{81}, 033413 (2010).

\bibitem{rosh4}
A. I. Voroshilo and S. P. Roshchupkin, Laser Phys. Lett. \textbf{2}, 184 (2005).


\bibitem{PanekMoller2004}
P. Panek, J. Z. Kami\'nski, and F. Ehlotzky, Phys. Rev. A \textbf{69}, 013404 (2004).

\bibitem{PanekMoller2004a}
P. Panek, J. Z. Kami\'nski, and F. Ehlotzky, Laser Phys. \textbf{14}, 1200 (2004).

\bibitem{AtomsFelipe2019}
F. Cajiao V\'elez, J. Z. Kami\'nski, and K. Krajewska, Atoms \textbf{7}, 34 (2019).

\bibitem{cep1}
K. Krajewska and J. Z. Kami\'nski, Phys. Rev. A \textbf{85}, 043404 (2012).

\bibitem{cep2}
S. Augustin and C. M\"uller, Phys. Rev. A \textbf{88}, 022109 (2013).

\bibitem{BW5}
M. J. A. Jansen and C. M\"uller, Phys. Rev. D \textbf{93}, 053011 (2016).

\bibitem{ItzyksonZuber}
C. Itzykson and J.-B. Zuber, \textit{Quantum Field Theory}, (McGraw-Hill, New York, 1980).

\bibitem{Boca2010}
M. Boca and V Florescu, Rom. Journ. Phys. \textbf{55}, 511 (2010).

\bibitem{Boca2011}
M. Boca, J. Phys. A: Math. Theor. \textbf{44}, 445303 (2011).

\bibitem{Antonino2018}
A. Di~Piazza, Phys. Rev. D \textbf{97}, 056028 (2018).

\bibitem{Wang2019}
H. Wang, M. Zhong, and L.-F. Gan, Commun. Theor. Phys. \textbf{71}, 1179 (2019).

\bibitem{Serov2017}
V. Serov, \textit{Fourier Series, Fourier Transform and Their Applications to Mathematical Physics}, (Springer International Publishing AG, Cham, Switzerland, 2017).

\bibitem{HHIM2012}
C. Harvey, T. Heinzl, A. Ilderton, and M. Marklund, Phys. Rev. Lett. \textbf{109}, 100402 (2012).

\bibitem{KK2012b}
K. Krajewska and J. Z. Kami\'nski, Phys. Rev. A \textbf{86}, 052104 (2012).

\bibitem{Reiss2014}
H. R. Reiss, Phys. Rev. A \textbf{89}, 022116 (2014).

\bibitem{KCK2015a}
K. Krajewska, F. Cajiao V\'elez, and J. Z. Kami\'nski, Phys. Rev. A \textbf{91}, 062106 (2015).

\bibitem{muon1}
S. J. M\"uller and C. M\"uller, Phys. Rev. D \textbf{80}, 053014 (2009).

\bibitem{muon2}
V. N. Nedoreshta, A. I. Voroshilo, and S. P. Roshchupkin, Eur. Phys. Jour. D \textbf{48},451 (2008).

\bibitem{muon3}
W.-Y Du, P.-F. Zhang, and B.-H. Wang, Front. Phys. \textbf{13}, 133401 (2018).

\bibitem{muon4}
N. Wang., L. Jiao, and A. Liu, Chin. Phys. B \textbf{28}, 193402 (2019).

\bibitem{muon5}
E. A. Padusenko, S, P. Roshchupkin, and A. I. Voroshilo, Laser Phys. Lett. \textbf{6}, 242 (2008).

\bibitem{Compton1}
M. Boca and V. Florescu, Phys. Rev. A \textbf{80}, 053403 (2009); {\it ibid.}, Phys. Rev. A \textbf{81}, 039901 (2010).

\bibitem{Compton2}
F. Mackenroth and A. Di Piazza, Phys. Rev. A \textbf{83}, 032106 (2011).

\bibitem{Compton3}
D. Seipt and B. K\"ampfer, Phys. Rev. A \textbf{83}, 022101 (2011).

\bibitem{Compton4}
D. Seipt, S. G. Rykovanov, A. Surzhykov, and S.~Fritzsche, Phys. Rev. A \textbf{91}, 033402 (2015).

\bibitem{Compton5}
K. Krajewska and J. Z. Kami\'nski, Phys. Rev. A \textbf{85}, 062102 (2012).

\bibitem{Compton6}
T. N. Wistisen, Phys. Rev. D \textbf{90}, 125008 (2014); {\it ibid.}, Phys. Rev. D \textbf{91}, 069903 (2015).

\bibitem{Compton7}
J. P. Corson and J. Peatross, Phys. Rev. A \textbf{84}, 053832 (2011).

\bibitem{BW1}
A. I. Titov, H. Takabe, B. K\"ampfer, and A. Hosaka, Phys. Rev. Lett. \textbf{108}, 240406 (2012).

\bibitem{BW2}
T. Nousch, D. Seipt, B. K\"ampfer, and A.I. Titov, Phys. Lett. B \textbf{715}, 246 (2012).

\bibitem{BW3}
M. J. Duff, R. Capdessus, C. P. Ridgers, and P.~McKenna, Plasma Phys. Control. Fusion \textbf{61}, 094001 (2019).

\bibitem{BW4}
M. Lobet, X. Davoine, E. d'Humi\`eres, and L. Gremillet, Phys. Rev. Accel. Beams 20, 043401 (2017).

\bibitem{Ehlotzky2001}
F. Ehlotzky, Phys. Rep. \textbf{345}, 175 (2001).

\bibitem{Shapiro2003}
M. Shapiro and P. Brumer, Rep. Prog. Phys. \textbf{66}, 859 (2003).

\bibitem{cep4}
J. Z. Kami\'nski and F. Ehlotzky, Phys. Rev. A \textbf{50}, 4404 (1994).

\bibitem{cep7}
S. Varr\'o and F. Ehlotzky, J. Phys. B \textbf{30}, 1061 (1997).

\bibitem{cep3}
S. P. Roshchupkin and A. A. Lebed', Phys. Rev. A \textbf{90}, 035403 (2014).

\bibitem{cep5}
M. M. Dellweg and C. M\"uller, Phys. Rev. A \textbf{91}, 062102 (2015).

\bibitem{cep6}
M. M. Dellweg, H. M. Awwad, and C. M\"uller, Phys. Rev. A \textbf{94}, 022122 (2016).

\bibitem{cepp1}
F. Mackenroth, A. Di Piazza, and C. H. Keitel, Phys. Rev. Lett. \textbf{105}, 063903 (2010).

\bibitem{cepp2}
K. Krajewska, M. Twardy, and J. Z. Kami\'nski, Phys. Rev. A \textbf{89}, 052123 (2014).

\bibitem{cepp3}
J.-X. Li, Y.-Y. Chen, K. Z. Hatsagortsyan, and C. H. Keitel, Phys. Rev. Lett. \textbf{120}, 124803 (2018).

\bibitem{Deeksha}
D. Kanti, J. Z. Kami\'nski, L.-Y. Peng, and K. Krajewska, Phys. Rev. A {\bf 104}, 033112 (2021).

\bibitem{res0}
M. L. Goldberger and K. M. Watson, \textit{Collision Theory}, (John Wiley \& Sons, New York, 1964).

\bibitem{res1}
V. I. Kukulin, V. M. Krasnopol'sky, and J. Hor\'a\v{c}ek, \textit{Theory of Resonances. Principles and Applications}, (Kluwer, Dordrecht, 1989).

\bibitem{res2}
R. G. Newton, \textit{Scattering Theory of Waves and Particles}, (Springer, New York, 1982).

\bibitem{cep8}
A.I. Titov, A. Otto, and B. K\"ampfer, Eur. Phys. J. D {\bf 74}, 39 (2020).

\bibitem{LorentzFano1}
J. Z. Kami\'nski, A. Jaro\'n, and F. Ehlotzky, J. Phys. B \textbf{28}, 4895 (1995).

\bibitem{LorentzFano2}
C. Ott, A. Kaldun, R. Raith, K. Meyer, M. Laux, J. Evers, C. H. Keitel, C. H. Greene, and T. Pfeifer, Science \textbf{340}, 716 (2013).

\bibitem{res3}
R. H. Dalitz, \textit{Resonance: Its description, criteria and significance}. In: S. Albeverio,  L. S. Ferreira, L. Streit (eds) \textit{Resonances - Models and Phenomena}. Lecture Notes in Physics \textbf{211} (Springer, Berlin, 1984).

\bibitem{res4}
J. R. Taylor, \textit{Scattering Theory. The Quantum Theory of Nonrelativistic Collisions}, (John Wiley \& Sons, New York, 1972).

\bibitem{Fano1}
U. Fano, Nuovo Cim. \textbf{12}, 154 (1935).

\bibitem{Fano2}
U. Fano, Phys. Rev. \textbf{124}, 1866 (1961).

\bibitem{BHRos}
S. P. Roshchupkin, N. R. Larin, and V. V. Dubov, Phys. Rev. D {\bf 104}, 116011 (2021).

\bibitem{bremRos}
S. P. Roshchupkin, A. V. Dubov, V. V. Dubov, and S. S. Starodub, New J. Phys. {\bf 24}, 013020 (2022).

\end{thebibliography}
\end{document}